\documentclass[a4paper,11pt]{article}
\pdfoutput=1 

\usepackage{jheppub} 
\makeatletter
\DeclareRobustCommand*{\bfseries}{%
  \not@math@alphabet\bfseries\mathbf
  \fontseries\bfdefault\selectfont
  \boldmath
}
\makeatother

\usepackage{calc}
\usepackage{rotating}
\usepackage[english]{babel}
\usepackage{graphicx}
\usepackage{subfig}
\usepackage{float}
\usepackage{amsmath}
\usepackage{amssymb}
\usepackage{amsthm}
\usepackage{latexsym}
\usepackage{dcolumn}
\newcommand{\newc}{\newcommand*}

\long\def\begincomment#1\endcomment{%
        \begingroup\sf\baselineskip12pt#1\endgroup}

\newc{\etal}{\textrm{et al.}} 
\newc{\eg}{\textrm{e.g.}} 
\newc{\ie}{\textrm{i.e.}}
\newc{\etc}{\textrm{etc}}
\newc\vs{\textrm{vs.}}
\newc{\cl}{\rm {C.L.}}

\newc{\ev}{\ensuremath{\,\mathrm{eV}}}
\newc{\kev}{\ensuremath{\,\mathrm{keV}}}
\newc{\mev}{\ensuremath{\,\mathrm{MeV}}}
\newc{\gev}{\ensuremath{\,\mathrm{GeV}}}
\newc{\tev}{\ensuremath{\,\mathrm{TeV}}}
\newc{\MeV}{\mev} 
\newc{\TeV}{\tev}
\newc{\invpb}{\ensuremath{/\text{pb}}}
\newc{\invfb}{\ensuremath{\,\text{fb}^{-1}}}
\newc\nb{\ensuremath{\,\mathrm{nb}}} \newc\pb{\ensuremath{\,\mathrm{pb}}} \newc\fb{\ensuremath{\,\mathrm{fb}}}
\newc\pc{\ensuremath{\,\mathrm{pc}}}
\newc\kpc{\ensuremath{\,\mathrm{kpc}}}
\newc\mpc{\ensuremath{\,\mathrm{Mpc}}}
\newc\ps{\ensuremath{\,\mathrm{ps}}} 
\newc\cmeter{\ensuremath{\,\mathrm{cm}}} 
\newc\meter{\ensuremath{\,\mathrm{m}}} 
\newc\kmeter{\ensuremath{\,\mathrm{km}}}
\newc\second{\ensuremath{\,\mathrm{s}}}
\newc\msecond{\ensuremath{\,\mathrm{ms}}}
\newc\nsecond{\ensuremath{\,\mathrm{ns}}}
\newc\psecond{\ensuremath{\,\mathrm{ps}}}

\newc{\chisqmin}{\ensuremath{\chi^2_{\mathrm{min}}}}
\newc{\Delchisq}{\ensuremath{\Delta\chi^2}}
\newc{\chisq}{\ensuremath{\chi^2}}
\newc{\like}{\ensuremath{\mathcal{L}}}

\newc\lsim{\ensuremath{\mathrel{\rlap{\lower4pt\hbox{\hskip1pt$\sim$}}\raise1pt\hbox{$<$}}}}
\newc\gsim{\ensuremath{\mathrel{\rlap{\lower4pt\hbox{\hskip1pt$\sim$}}\raise1pt\hbox{$>$}}}}
\newc{\VEV}[1]{\ensuremath{\langle #1 \rangle}}
\newc{\dl}{\ensuremath{\stackrel{\leftarrow}{D}}}
\newc{\dr}{\ensuremath{\stackrel{\rightarrow}{D}}}

\newc{\bcenter}{\begin{center}}    \newc{\ecenter}{\end{center}}
\newc{\bfl}{\begin{flushleft}}    \newc{\efl}{\end{flushleft}}
\newc{\bfr}{\begin{flushright}}    \newc{\efr}{\end{flushright}}

\newc{\bi}{\begin{itemize}}
\newc{\ei}{\end{itemize}}
\newc{\bed}{\begin{description}}
\newc{\eed}{\end{description}}
\newc{\ben}{\begin{enumerate}}
\newc{\een}{\end{enumerate}}

\newc{\be}{\begin{equation}}
\newc{\ee}{\end{equation}}
\newc{\bea}{\begin{eqnarray}}
\newc{\eea}{\end{eqnarray}}
\newc{\ra}{\rightarrow}

\newc{\alphas}{\ensuremath{\alpha_s}}
\newc{\alphatwo}{\ensuremath{\alpha_2}}
\newc{\alphaone}{\ensuremath{\alpha_1}}
\newc{\alphai}[1]{\ensuremath{\alpha_{#1}}}
\newc{\alphaem}{\ensuremath{\alpha_{\mathrm{em}}}}
\newc{\alphaeff}{\ensuremath{\alpha_{\mathrm{eff}}}}
\newc{\sineff}{\ensuremath{\sin \theta_{\mathrm{eff}}}}
\newc{\sinsqeff}{\ensuremath{\sin^2 \theta_{\mathrm{eff}}}}
\newc{\dalphahad}{\ensuremath{\Delta \alpha_{\mathrm{had}}}}
\newc{\yt}{\ensuremath{h_t}} \newc{\yb}{\ensuremath{h_b}} \newc{\ytau}{\ensuremath{h_{\tau}}}
\newc\mz{\ensuremath{M_Z}} 
\newc\mw{\ensuremath{m_W}}
\newc\mZ{\mz}        \newc\mW{\mw}
\newc\mhsm{\ensuremath{ m_{H_{\mathrm{SM}}}}}
\newc{\mtop}{\ensuremath{ m_t}}               \newc{\mtpole}{\ensuremath{ M_t}}
\newc{\mbottom}{\ensuremath{ m_b}} 
\newc{\mtau}{\ensuremath{ m_{\tau}}}
\newc{\mt}{\mtpole}
\newc{\mb}{\mbottom} 
\newc{\rgg}{\ensuremath{R_{h}(\gamma\gamma)}}
\newc{\rzz}{\ensuremath{R_{h}(ZZ)}}
\newc{\rtwogg}{\ensuremath{R_{h_2}(\gamma\gamma)}}
\newc{\rtwozz}{\ensuremath{R_{h_2}(ZZ)}}
\newc{\ronegg}{\ensuremath{R_{h_1}(\gamma\gamma)}}
\newc{\ronezz}{\ensuremath{R_{h_1}(ZZ)}}
\newc{\rsiggg}{\ensuremath{R_{h_\textrm{sig}}(\gamma\gamma)}}
\newc{\rsigzz}{\ensuremath{R_{h_\textrm{sig}}(ZZ)}}
\newc{\llbar}{\ensuremath{\ell\bar{\ell}}}
\newc{\tauptaum}{\ensuremath{ \tau^+\tau^-}}
\newc{\qqbar}{\ensuremath{ q\bar{q}}} \newc{\ppbar}{\ensuremath{ p\bar{p}}}
\newc{\bbbar}{\ensuremath{ b\bar{b}}} \newc{\ttbar}{\ensuremath{ t\bar{t}}}
\newc{\ffbar}{\ensuremath{ f\bar{f}}} \newc{\tautaubar}{\ensuremath{ \tau\bar{\tau}}}
\newc{\mchi}{\ensuremath{m_{\chi}}}
\newc{\squark}{\ensuremath{\tilde{q}}}
\newc{\slepton}{\ensuremath{\tilde{l}}}
\newc{\gluino}{\ensuremath{\tilde{g}}} 
\newc{\mgluino}{\ensuremath{{m_{\gluino}}}}
\newc{\tone}{\ensuremath{{\tilde{t}_1}}}

\newc{\sthw}{\ensuremath{ \sin\theta_W}}              \newc{\cthw}{\ensuremath{\cos\theta_W}}
\newc{\tanthw}{\ensuremath{ \tan\theta_W}}              \newc{\cotthw}{\ensuremath{\cot\theta_W}}
\newc{\ssqthw}{\ensuremath{\sin^2 \theta_W}}
\newc{\msbar}{\ensuremath{\overline{MS}}} \newc{\drbar}{\ensuremath{\overline{DR}}}
\newc{\mtmtsmmsbar}{\ensuremath{ m_t(m_t)^{\msbar}_{{\mathrm{SM}}}}}
\newc{\mtmtsmdrbar}{\ensuremath{ m_t(m_t)^{\drbar}_{{\mathrm{SM}}}}}
\newc{\mtmtmssmdrbar}{\ensuremath{ m_t(m_t)^{\drbar}_{{\mathrm{SUSY}}}}}
\newc{\mbmbmsbar}{\ensuremath{ m_b(m_b)^{\msbar} }}
\newc{\mbmbsmmsbar}{\ensuremath{ m_b(m_b)^{\msbar}_{{\mathrm{SM}}}}}
\newc{\mbmzsmmsbar}{\ensuremath{ m_b(\mz)^{\msbar}_{{\mathrm{SM}}}}}
\newc{\mbmzsmdrbar}{\ensuremath{ m_b(\mz)^{\drbar}_{{\mathrm{SM}}}}}
\newc{\mbmzmssmdrbar}{\ensuremath{ m_b(\mz)^{\drbar}_{{\mathrm{SUSY}}}}}
\newc{\mtaumzsmmsbar}{\ensuremath{ m_{\tau}(\mz)^{\msbar}_{{\mathrm{SM}}}}}
\newc{\mtaumzsmdrbar}{\ensuremath{ m_{\tau}(\mz)^{\drbar}_{{\mathrm{SM}}}}}
\newc{\mtaumzmssmdrbar}{\ensuremath{ m_{\tau}(\mz)^{\drbar}_{{\mathrm{SUSY}}}}}
\newc{\alphasmzms}{\ensuremath{\alpha_s(M_Z)^{\overline{MS}}}}
\newc{\alphaimzms}[1]{\ensuremath{\alpha_{#1}(M_Z)^{\overline{MS}}}}
\newc{\alphaemmz}{\ensuremath{\alpha_{\mathrm{em}}(M_Z)^{\overline{MS}}}}

\newc{\mzero}{\ensuremath{{m_0}}}
\newc{\mhalf}{\ensuremath{ m_{1/2}}}
\newc{\tanb}{\ensuremath{\tan\beta}}
\newc{\azero}{\ensuremath{ A_0}}
\newc{\bzero}{\ensuremath{ B_0}}
\newc{\signmu}{\ensuremath{\rm{sgn}\,\mu}}
\newc{\mueff}{\ensuremath{\mu_{\rm{eff}}}}
\newc{\lam}{\ensuremath{{\lambda}}}
\newc{\kap}{\ensuremath{{\kappa}}}
\newc{\alam}{\ensuremath{{A_{\lambda}}}}
\newc{\akap}{\ensuremath{{A_{\kappa}}}}
\newc{\hs}{\ensuremath{ H_s}}      
\newc{\mhs}{\ensuremath{ m_{H_s}}} 
\newc{\mgut}{\ensuremath{ M_{\rm GUT}}}
\newc{\mplanck}{\ensuremath{ M_{\rm P}}}      \newc{\mpl}{\ensuremath{ M_{\rm Pl}}}
\newc{\msusy}{\ensuremath{ M_{\rm SUSY}}}      \newc{\ms}{\ensuremath{ M_{\rm S}}}
 \newc{\mhl}{\ensuremath{m_\hl}} 
 \newc{\mhone}{\ensuremath{m_{h_1}}} 
 \newc{\mhtwo}{\ensuremath{m_{h_2}}} 
 \newc{\mglu}{\ensuremath{m_{\tilde g}}} 
 \newc{\mul}{\ensuremath{m_{\tilde{u}_L}}} 
 \newc{\mtone}{\ensuremath{m_{\tilde{t}_1}}} 
 \newc{\ma}{\ensuremath{m_A}} 
 \newc{\maone}{\ensuremath{m_{a_1}}} 
 \newc{\matwo}{\ensuremath{m_{a_2}}}
 \newc{\hone}{\ensuremath{h_1}}
 \newc{\htwo}{\ensuremath{h_2}}
 \newc{\aone}{\ensuremath{a_1}}
 \newc{\atwo}{\ensuremath{a_2}}
 \newc{\mhu}{\ensuremath{ m_{H_u}}}       
 \newc{\mhd}{\ensuremath{ m_{H_d}}}
 \newc{\mhusq}{\ensuremath{ m_{H_u}^2}}       
 \newc{\mhdsq}{\ensuremath{ m_{H_d}^2}}
 \newc{\mhuew}{\ensuremath{ m^{\ast}_{H_u}}}       
 \newc{\mhdew}{\ensuremath{ m^{\ast}_{H_d}}}
 \newc{\mhuewsq}{\ensuremath{ m^{\ast\, 2}_{H_u}}}       
 \newc{\mhdewsq}{\ensuremath{ m^{\ast\, 2}_{H_d}}}
 \newc{\hu}{\ensuremath{ H_u}}       
 \newc{\hd}{\ensuremath{ H_d}}
 \newc{\barmhu}{\ensuremath{ \bar{m}_{H_u}}}
 \newc{\barmhd}{\ensuremath{ \bar{m}_{H_d}}}

 \newc{\mqthree}{\ensuremath{m_{\widetilde{Q}_3}^2}}
 \newc{\muthree}{\ensuremath{m_{\tilde{u}_3}^2}}
 \newc{\mdthree}{\ensuremath{m_{\tilde{d}_3}^2}}
 \newc{\mlthree}{\ensuremath{m_{\widetilde{L}_3}^2}}
 \newc{\methree}{\ensuremath{m_{\tilde{e}_3}^2}}
 \newc{\mqtwo}{\ensuremath{m_{\widetilde{Q}_2}^2}}
 \newc{\mutwo}{\ensuremath{m_{\tilde{u}_2}^2}}
 \newc{\mdtwo}{\ensuremath{m_{\tilde{d}_2}^2}}
 \newc{\mltwo}{\ensuremath{m_{\widetilde{L}_2}^2}}
 \newc{\metwo}{\ensuremath{m_{\tilde{e}_2}^2}}
 \newc{\mqone}{\ensuremath{m_{\widetilde{Q}_1}^2}}
 \newc{\muone}{\ensuremath{m_{\tilde{u}_1}^2}}
 \newc{\mdone}{\ensuremath{m_{\tilde{d}_1}^2}}
 \newc{\mlone}{\ensuremath{m_{\widetilde{L}_1}^2}}
 \newc{\meone}{\ensuremath{m_{\tilde{e}_1}^2}}
 \newc{\mone}{\ensuremath{M_1}}
 \newc{\monesq}{\ensuremath{M_1^2}}
 \newc{\mtwo}{\ensuremath{M_2}}
 \newc{\mtwosq}{\ensuremath{M_2^2}}
 \newc{\mthree}{\ensuremath{M_3}}
 \newc{\mthreesq}{\ensuremath{M_3^2}}
 \newc{\atau}{\ensuremath{{A_{\tau}}}}
 \newc{\at}{\ensuremath{{A_{t}}}}
 \newc{\ab}{\ensuremath{{A_{b}}}}
 \newc{\atausq}{\ensuremath{{A_{\tau}^2}}}
 \newc{\atsq}{\ensuremath{{A_{t}^2}}}
 \newc{\absq}{\ensuremath{{A_{b}^2}}}

 \newc{\dmzero}{\ensuremath{\Delta{_{m_0}}}}
 \newc{\dmhalf}{\ensuremath{\Delta{_{m_{1/2}}}}}
 \newc{\dmu}{\ensuremath{\Delta{_{\mu}}}}

 \newc{\pten}{\ensuremath{\psi_{10}}}
 \newc{\ffive}{\ensuremath{\phi_{5}}}
 \newc{\hfive}{\ensuremath{h_{5}}}
 \newc{\hbfive}{\ensuremath{h_{\bar{5}}}}
 \newc{\thet}{\ensuremath{\theta_{50}}}
 \newc{\thetb}{\ensuremath{\theta_{\,\overline{50}}}}
 \newc{\ptenhat}{\ensuremath{\hat{\psi}_{10}}}
 \newc{\ffivehat}{\ensuremath{\hat{\phi}_{5}}}
 \newc{\hfivehat}{\ensuremath{\hat{h}_{5}}}
 \newc{\hbfivehat}{\ensuremath{\hat{h}_{\bar{5}}}}
 \newc{\thethat}{\ensuremath{\hat{\theta}_{50}}}
 \newc{\thetbhat}{\ensuremath{\hat{\theta}_{\,\overline{50}}}}
 \newc{\si}{\ensuremath{\Sigma}}
 \newc{\mfive}{\ensuremath{m_5^2}}
 \newc{\mten}{\ensuremath{m_{10}^2}}
 \newc{\dfive}{\ensuremath{\Delta^2_5}}
 \newc{\dbfive}{\ensuremath{\Delta^2_{\bar{5}}}}
 \newc{\dfifty}{\ensuremath{\Delta^2_{50}}}
 \newc{\dfiftyb}{\ensuremath{\Delta^2_{\,\overline{50}}}}
 \newc{\msi}{\ensuremath{m_{\Sigma}^2}}
 \newc{\lamh}{\ensuremath{\lambda_{H}}}
 \newc{\lamhb}{\ensuremath{\lambda_{\bar{H}}}}
 \newc{\ah}{\ensuremath{A_{H}}}
 \newc{\ahb}{\ensuremath{A_{\bar{H}}}}
 \newc{\lams}{\ensuremath{\lambda_{S}}}
 \newc{\as}{\ensuremath{A_{S}}}
 \newc{\lamsig}{\ensuremath{\lambda_{\si}}}
 \newc{\asig}{\ensuremath{A_{\si}}}

 \newc{\msten}{\ensuremath{m_{16}^2}}
 \newc{\mhun}{\ensuremath{m_{126}^2}}
 \newc{\mhunb}{\ensuremath{m_{\bar{126}}^2}}
 \newc{\mthun}{\ensuremath{m_{210}^2}}
 \newc{\ahun}{\ensuremath{A_{\bar{126}}}}
 \newc{\yhun}{\ensuremath{Y_{\bar{126}}}}
 \newc{\aten}{\ensuremath{A_{10}}}
 \newc{\yten}{\ensuremath{Y_{10}}}
 \newc{\alone}{\ensuremath{A_{\lambda_1}}}
 \newc{\altwo}{\ensuremath{A_{\lambda_2}}}
 \newc{\althree}{\ensuremath{A_{\lambda_3}}}
 \newc{\althreeb}{\ensuremath{A_{\bar{\lambda_3}}}}
 \newc{\lone}{\ensuremath{\lambda_1}}
 \newc{\ltwo}{\ensuremath{\lambda_2}}
 \newc{\lthree}{\ensuremath{\lambda_3}}
 \newc{\lthreeb}{\ensuremath{\bar{\lambda_3}}}

\newc{\sigsip}{\ensuremath{\sigma^{\rm SI}_{p}}}	\newc{\sigsin}{\ensuremath{\sigma^{\rm SI}_{n}}}
\newc{\sigsdp}{\ensuremath{\sigma^{\rm SD}_{p}}}	\newc{\sigsdn}{\ensuremath{\sigma^{\rm SD}_{n}}}
\newc{\sigsi}{\ensuremath{\sigma^{\rm SI}}}	\newc{\sigsd}{\ensuremath{\sigma^{\rm SD}}}
\newc{\sigv}{\ensuremath{\sigma v}}
\newc{\abund}{\ensuremath{ \Omega h^2}}
\newc{\omegadm}{\ensuremath{ \Omega_{{\rm DM}}}}     \newc{\abunddm}{\ensuremath{ \Omega_{{\rm DM}} h^2}} 
\newc{\omegam}{\ensuremath{ \Omega_{{\rm m}}}}       \newc{\abundm}{\ensuremath{ \Omega_{{\rm m}} h^2}}
\newc{\omegab}{\ensuremath{ \Omega_{{\rm b}}}}	\newc{\abundb}{\ensuremath{ \Omega_{{\rm b}} h^2}}
\newc{\omegatot}{\ensuremath{ \Omega_{{\rm TOT}}}}
\newc{\omegacdm}{\ensuremath{ \Omega_{{\rm CDM}}}}   \newc{\abundcdm}{\ensuremath{ \Omega_{{\rm CDM}} h^2}}
\newc{\omegalambda}{\ensuremath{ \Omega_{\Lambda}}} \newc{\abundlambda}{\ensuremath{ \Omega_{\Lambda} h^2}}
\newc{\omegarad}{\ensuremath{ \Omega_{{\rm rad}}}}  \newc{\abundrad}{\ensuremath{ \Omega_{{\rm rad}} h^2}}
\newc{\rhocrit}{\ensuremath{ \rho_{\rm crit}}}
\newc{\rhochi}{\ensuremath{ \rho_{\chi}}}
\newc{\abunchi}{\ensuremath{\Omega_\chi h^2}}
\newc{\abundlsp}{\ensuremath{\Omega_{\rm LSP}h^2}}
\newcommand*{\abundchi}{\ensuremath{\Omega_\chi h^2}}

\newc{\amu}{\ensuremath{ a_{\mu}}}        \newc{\amususy}{\ensuremath{ a_{\mu}^{\mathrm{SUSY}}}}
\newc{\amuexpt}{\ensuremath{ a_{\mu}^{\mathrm{expt}}}}        \newc{\amusm}{\ensuremath{ a_{\mu}^{\mathrm{SM}}}}
\newc\deltaamu{\ensuremath{\Delta a_{\mu}}} \newc{\deltaamususy}{\ensuremath{\delta a_{\mu}^{\mathrm{SUSY}}}}
\newc\gmtwo{\ensuremath{ (g-2)_{\mu}}} 
\newc{\deltagmtwomususy}{\ensuremath{\delta\left(g-2\right)_{\mu}^{\mathrm{SUSY}}}}
\newc{\deltagmtwomu}{\ensuremath{\delta\left(g-2\right)_{\mu}}}
\newc\BR{\ensuremath{\rm BR}}
\newc\bsgamma{\ensuremath{ b\rightarrow s \gamma }}
\newc\bxsgamma{\ensuremath{\overline{B}\rightarrow X_{s}\gamma}}
\newc\brbsgamma{\ensuremath{\BR\left(\bsgamma\right)}}
\newc\brbxsgamma{\ensuremath{\BR\left(\bxsgamma\right)}}
\newc\bsmumu{\ensuremath{B_s\to\mu^+\mu^-}}
\newc\brbsmumu{\ensuremath{\BR\left(B_s\to\mu^+\mu^-\right)}}
\newc\bdmmumu{\ensuremath{\overline{B}_d\to\mu^+\mu^-}}
\newc\bbbarmix{\ensuremath{\overline{B}_s\mbox{-}B_s}}      
\newc\delmbs{\ensuremath{\Delta M_{B_s}}}
\newc{\butaunu}{\ensuremath{B_u \rightarrow \tau \nu}}
\newc{\brbutaunu}{\ensuremath{\BR\left(B_u \rightarrow \tau \nu\right)}}

\newcommand*{\reffig}[1]{Fig.~\ref{#1}}

     \newcommand*{\refsec}[1]{Sec.~\ref{#1}}

\newcommand*{\neuttwo}{\ensuremath{\tilde{{\chi}}^0_2}}

\newcommand*{\charone}{\ensuremath{\chi^{\pm}_1}}

\newcommand*{\stau}{\ensuremath{\tilde{\tau}}}

\newcommand*{\mstopone}{\ensuremath{m_{\tilde{t}_1}}}
\newcommand*{\mstoptwo}{\ensuremath{m_{\tilde{t}_2}}}

\newcommand*{\xenononet}{\text{XENON-1T}}

\let\oldcite\cite
\renewcommand*{\cite}{~\oldcite}

\newcommand*{\hl}{\ensuremath{h}}

\restylefloat{figure}

\title{Prospects for dark matter searches in the pMSSM}

\author[1]{Leszek Roszkowski,\note{On leave of absence from the University of Sheffield, U.K.}}
\author{Enrico Maria Sessolo}
\author{and Andrew J.~Williams}

\affiliation{National Centre for Nuclear Research,\\
  Ho{\. z}a 69, 00-681 Warsaw, Poland} 

\emailAdd{L.Roszkowski@sheffield.ac.uk}
\emailAdd{Enrico-Maria.Sessolo@fuw.edu.pl}
\emailAdd{Andrew.Williams@fuw.edu.pl}

\abstract{We investigate the prospects for detection of neutralino dark matter
  in the 19-parameter phenomenological MSSM (pMSSM).  We explore very
  wide ranges of the pMSSM parameters but pay particular attention to
  the higgsino-like neutralino at the $\sim1\tev$ scale, which has
  been shown to be a well motivated solution in many constrained
  supersymmetric models, as well as to a wino-dominated solution with
  the mass in the range of 2--3\tev.  After summarising the present
  bounds on the parameter space from direct and indirect detection
  experiments, we focus on prospects for detection of the Cherenkov
  Telescope Array (CTA). To this end, we derive a realistic assessment
  of the sensitivity of CTA to photon fluxes from dark matter
  annihilation by means of a binned likelihood analysis for the
  Einasto and Navarro-Frenk-White halo profiles.  We use the most up
  to date instrument response functions and background simulation
  model provided by the CTA Collaboration. We find that, with 500
  hours of observation, under the Einasto profile CTA is bound to
  exclude at the 95\%~C.L. almost all of the $\sim 1\tev$ higgsino
  region of the pMSSM, effectively closing the window for heavy
  supersymmetric dark matter in many realistic models. CTA will be
  able to probe the vast majority of cases corresponding
  to a spin-independent scattering cross section
  below the reach of 1-tonne underground detector searches for dark
  matter, in fact even well below the irreducible neutrino background for
  direct detection. On the other hand, many points lying beyond the
  sensitivity of CTA will be within the reach of 1-tonne 
  detectors, and some within collider reach. Altogether, CTA will
  provide a highly sensitive way of searching for dark matter that
  will be partially overlapping and partially complementary with
  1-tonne detector and collider searches, thus being
  instrumental to effectively explore the nearly full parameter space of
  the pMSSM.}

\begin{document}
\maketitle
\flushbottom

\section{\label{sec:intro}Introduction}

The search for particles that comprise the dark matter (DM) in the Universe
has in recent years made much progress. Alternative and complementary experimental strategies 
are employed ranging from direct detection of DM-nucleon scattering in underground laboratories, to 
indirect detection of DM through observation of the Standard Model (SM) products of annihilation in astrophysical phenomena
(for a recent review of the large number of experiments 
dedicated to direct and indirect detection of DM see, e.g., the updated version of\cite{Drees:2012ji} in\cite{Agashe:2014kda} and References therein),
to DM direct production at colliders\cite{Goodman:2010ku,Fox:2011pm}.
  
The most impressive advances in sensitivity
have arguably been made in direct detection experiments, where improvement 
happened rapidly and led to the most recent null results by XENON100\cite{Aprile:2012nq} 
and LUX\cite{Akerib:2013tjd}. As a consequence, the upper bounds on the spin-independent DM-nucleon elastic scattering cross section,
\sigsip, have become increasingly constraining for many models of weakly interacting massive particles (WIMPs).
On the other hand, interesting upper bounds on the cross section for WIMP production\cite{Aad:2013oja,Aad:2014vka,Khachatryan:2014rra,CMS-PAS-EXO-12-047}
have been placed at the LHC, which have become particularly constraining for many models of low-mass DM. 
At the same time, strong limits on the present-day DM annihilation cross section, \sigv, as a function of the 
WIMP mass, have been provided by $\gamma$-ray experiments. In particular, the most stringent ones for masses up to $\sim1\tev$
come from Fermi-LAT's data on dwarf Spheroidal Galaxies (dSphs)\cite{Ackermann:2013yva}. 
For larger masses the air Cherenkov radiation telescope H.E.S.S. produces the strongest limits from observation of the Galactic Center (GC)\cite{HESS}. 
The strongest indirect limits on the spin-dependent DM-proton cross section, \sigsdp, 
have instead been obtained at IceCube/DeepCore\cite{IceCube:2011aj,Aartsen:2012kia} and ANTARES\cite{Adrian-Martinez:2013ayv} in observations of neutrinos from the Sun, 
and in monojet searches at the LHC\cite{Khachatryan:2014rra} 
for some choices of interactions and mediator masses. 

From a theoretical perspective, the most interesting solution to the DM puzzle arguably still 
comes from low scale supersymmetry (SUSY), as SUSY has the ability to solve 
many long-standing theoretical issues within one and the same elegant framework.
In common scenarios where the lightest SUSY particle (LSP) is the lightest neutralino of the Minimal Supersymmetric Standard Model (MSSM)
and makes up all of the DM in the Universe many of the experiments mentioned above 
have already started to exclude important parts of the parameter space. Some of the solutions\cite{Chan:1997bi,Feng:1999mn,Feng:1999zg} 
previously favoured by considerations of electro-weak (EW) naturalness, 
featuring the neutralino as a somewhat balanced admixture of higgsino and gaugino quantum states 
with a mass \mchi\ in the range 80 to 200\gev\
now show significant tension\cite{Fowlie:2013oua,Roszkowski:2014wqa} with the limits from XENON100 and LUX.
The same mixed solutions also start to show some tension\cite{Fowlie:2013oua} 
with the limits on \sigsdp\ from IceCube and ANTARES.
On the other hand, under the assumption that the lightest chargino and second lightest neutralino 
are not much heavier than the lightest neutralino, 
the regions of SUSY parameter space characterised by bino-like neutralinos with mass $\mchi\lesssim 100-300\gev$ 
(depending on whether light sleptons are present) are starting to be probed\cite{Fowlie:2013oua,Calibbi:2013poa,Calibbi:2014lga,Arbey:2013iza,Chakraborti:2014gea} 
by EW-ino searches at the LHC\cite{Aad:2014vma,Aad:2014nua,Khachatryan:2014qwa,Khachatryan:2014mma}.

At the same time, it has been
shown\cite{Strege:2012bt,Cabrera:2012vu,Kowalska:2013hha,Buchmueller:2013rsa,Roszkowski:2014wqa,Buchmueller:2014yva,Bechtle:2014yna} 
in global fits of the Constrained MSSM (CMSSM)\cite{Kane:1993td} and Non-Universal Higgs Model (NUHM) 
that one of the consequences of the discovery at the LHC of a $\sim 125\gev$ Higgs boson\cite{Aad:2012tfa,Chatrchyan:2012ufa} in agreement with the SM
is that the now favoured parameter space quite naturally gives rise to DM candidates heavier than previously thought.
Typically, the $\sim1\tev$ almost pure higgsino, whose existence as a solution for DM in the MSSM has been long known\cite{Profumo:2004at,ArkaniHamed:2006mb}
but in the framework of unified SUSY was first pointed out in a pre-LHC study of the NUHM\cite{Roszkowski:2009sm},
is now substantially favoured in the above mentioned CMSSM and NUHM and in some non-universal
scenarios characterised by reduced EW fine tuning\cite{Kaminska:2013mya,Kowalska:2014hza,Chakraborti}.
On the other hand, in ``split" SUSY scenarios with anomaly mediation\cite{Giudice:1998xp,Randall:1998uk}, 
which also rose in popularity\cite{Hall:2011jd,Arvanitaki:2012ps,Hall:2012zp,ArkaniHamed:2012gw} after the Higgs discovery, the wino is the DM. To be a thermal relic that satisfies the relic density 
its mass should be even larger, in the range 2--3\tev.

Incidentally, because of its relatively large cross sections for annihilations to SM particles, 
the wino LSP features excellent indirect detection prospects, 
which have been recently investigated in several papers\cite{Cohen:2013ama,Fan:2013faa,Hryczuk:2014hpa}.
The annihilation cross section is enhanced in this case by inclusion 
of the Sommerfeld enhancement\cite{Hisano:2002fk,Hisano:2004ds},
a nonperturbative effect which affects \sigv\ and also modifies the expectations for the relic density\cite{Hisano:2006nn,Cirelli:2007xd,Hryczuk:2010zi}. 
It is particularly substantial for the thermal wino, which has now consequently 
been excluded at the 95\%~C.L. by the absence of 
specific signatures in existing $\gamma$-ray observatories, particularly a monochromatic
line in H.E.S.S. data\cite{Abramowski:2013ax}. (The limit is relaxed\cite{Hryczuk:2014hpa} if one assumes a flat halo profile, like the Burkert profile\cite{Burkert:1995yz}.)
 
On the other hand, no such statement can be made for the other TeV-scale DM candidates
of the MSSM, like the $\sim1\tev$ almost pure higgsino for which the Sommerfeld enhancement is much less important.
Additionally, TeV-scale neutralinos lie probably outside of the direct reach of the LHC, 
and existing indirect detection experiments will not reach enough sensivity to test them either. 
The prospects for direct detection at 1-tonne detectors are, on the other hand, very good\cite{Fowlie:2013oua,Roszkowski:2014wqa} but 
upon hypothetical direct detection of a $1\tev$ higgsino 
complementary detection by some other means will be necessary to specify its properties. 

Such complementarity will likely be provided by the Cherenkov Telescope Array (CTA)\cite{Acharya:2013sxa}. 
The CTA project will build the next 
generation air Cherenkov telescope observatory. 
Several sensitivity studies\cite{Doro:2012xx,Wood:2013taa,Pierre:2014tra,Silverwood:2014yza} 
have shown that for DM masses greater than $\sim 100\gev$ CTA is expected to significantly exceed 
current limits for WIMP annihilation from the Cherenkov imaging telescopes
H.E.S.S.\cite{Abazajian:2011ak}, MAGIC\cite{Aleksic:2013xea}, and VERITAS\cite{AlexGeringer:2013fra}, and those from Fermi-LAT\cite{Ackermann:2013yva}.
CTA may even probe cross sections below the ``canonical" thermal relic
value of $2.6\times 10^{-26}\cmeter^3/\textrm{s}$ for some final states.
 
In a previous study\cite{Roszkowski:2014wqa} we showed that CTA has the potential to probe the $\sim1\tev$
higgsino region of the CMSSM and NUHM parameter space. As mentioned above, this is the region that our Bayesian analysis found to be 
favoured by the constraints in those models, 
encompassing approximately 70\% of the $2\sigma$ credible region in the CMSSM and 90\% of it in the NUHM.
In this paper we extend the study of direct and indirect DM detection prospects  
to the more general 19 dimensional low-scale parametrisation
called p19MSSM, or more commonly pMSSM\cite{Djouadi:1998di}. 
The reason for this is twofold: on the one hand, free gaugino mass terms 
will allow us to cover a greater number of possibilities for heavy SUSY DM: nearly pure higgsinos, winos, and bino/higgsino/wino admixtures; 
on the other hand, by additionally floating the scalar soft terms 
we try to fully incorporate the effects of the most common mechanisms 
of coannihilation with the lightest neutralino, and of mass degeneracies in general, on the calculation of the relic density.
As will be made clear in \refsec{sec:pointchar}, these effects can lead to substantial extensions of the allowed parameter space.
   
In this paper we will pay particular attention to neutralinos around the TeV scale, which seem to be favoured after the discovery of the Higgs boson,
and to the sensitivity of CTA, which is naturally poised to probe that particular region of the parameter space.
We also review the present status of direct and indirect detection constraints on the pMSSM
and compare the reach of CTA with that of 1-tonne direct detection experiments and of other detection methods.

The sensitivity of CTA to the pMSSM and its complementarity to other DM searches 
has also been recently analysed in Ref.\cite{Cahill-Rowley:2014boa}.
Some of the conclusions of this paper overlap with that study, but we also present several elements not included in\cite{Cahill-Rowley:2014boa}:\medskip   

$\bullet$ We use the most up to date\cite{JCarr} instrument response functions and background estimates provided by the CTA Collaboration\cite{montecarlo}.\medskip  

$\bullet$ The sensitivity of CTA is calculated from a binned likelihood function defined on the signal and background regions,
similarly to what was done in Ref.\cite{Silverwood:2014yza}. (The details of our calculation are presented at the end of this paper, in Appendix~\ref{sec:app}.)\medskip  

$\bullet$ We present results for CTA sensitivity under both the Einasto\cite{Einasto} and Navarro-Frenk-White (NFW)\cite{Navarro:1995iw} DM halo profiles.\medskip 

$\bullet$ Our analysis takes into account the Sommerfeld enhancement and the consequent limits on wino DM.\medskip  

The paper is organised as follows: in \refsec{sec:methodology} we specify the parameter and prior ranges of our scans and their distributions. 
We present there the set of experimental constraints applied to the likelihood. In \refsec{sec:pointchar} we describe the DM properties
of the parameter space regions favoured by the experimental constraints, the dominant annihilation mechanisms, and we 
identify a few benchmark points for future detection of DM. In \refsec{sec:results} we present a summary of the present status 
of direct and indirect bounds on neutralino DM and we expose future prospects for detection, 
particularly at CTA, for which we accurately calculate the sensitivity reach; we also 
present a comparison with present and future complementary experiments. We finally give our conclusions in \refsec{sec:summary}.
The details of our calculation of CTA's sensitivity including a treatment of alternative statistical approaches are given in Appendix~\ref{sec:app}.

\section{\label{sec:methodology}Scanning methodology and experimental constraints}

The pMSSM with 19 free parameters gives a generic 
coverage of the properties of the $CP$ and $R$ parity conserving MSSM. 
The parameters are defined at the scale of the geometrical average of the physical stop masses, 
$\msusy=(\mstopone\mstoptwo)^{1/2}$, and we scan them in the ranges given in Table~\ref{table:MSSMparams}. 
In addition, we scan over the top quark pole mass, $M_t$, 
treated here as a nuisance parameter. We assume a Gaussian distribution for $M_t$, 
whose central value and experimental error are given in\cite{ATLAS:2014wva},
$M_t=173.34\pm0.76\gev$. The remaining SM nuisance parameters are fixed to their PDG\cite{Agashe:2014kda} central values
as their variation is less relevant in our study. 

\begin{table}[t]
\begin{center}
\begin{tabular}{|c|c|}
\hline
\hline
Parameter & Range \\
\hline\hline
Higgsino/Higgs mass parameter & $-10\leq \mu\leq 10$ \\
Bino soft mass & $-10 \leq M_1 \leq10$ \\
Wino soft mass & $0.1\leq M_2 \leq 10$ \\
Gluino soft mass $^{\ast}$ &  $-10 \leq M_3 \leq 10$ \\
Top trilinear soft coupl. &$-10\leq A_{t}\leq 10 $\\
Bottom trilinear soft coupl. &$-10\leq A_{b}\leq 10 $\\
$\tau$ trilinear soft coupl. &$-10\leq A_{\tau}\leq 10 $\\
Pseudoscalar physical mass & $0.1 \leq m_A\leq 10$ \\
1st/2nd gen. soft L-slepton mass & $0.1\leq m_{\tilde{L}_1}\leq 10 $ \\
1st/2nd gen. soft R-slepton mass & $0.1\leq m_{\tilde{e}_R}\leq 10 $ \\
3rd gen. soft L-slepton mass & $0.1\leq m_{\tilde{L}_3}\leq 10$ \\
3rd gen. soft R-slepton mass & $0.1\leq m_{\tilde{\tau}_R}\leq 10$ \\
1st/2nd gen. soft L-squark mass & $0.75\leq m_{\tilde{Q}_1}\leq 10 $ \\
1st/2nd gen. soft R-squark up mass & $0.75\leq m_{\tilde{u}_R}\leq 10 $ \\
1st/2nd gen. soft R-squark down mass & $0.75\leq m_{\tilde{d}_R}\leq 10 $ \\
3rd gen. soft L-squark mass & $0.1\leq m_{\tilde{Q}_3}\leq 10$ \\
3rd gen. soft R-squark up mass & $0.1\leq m_{\tilde{t}_R}\leq 10 $ \\
3rd gen. soft R-squark down mass & $0.1\leq m_{\tilde{b}_R}\leq 10 $ \\
ratio of Higgs doublet VEVs & $1\leq \tan\beta\leq 62$ \\
\hline
\hline
\end{tabular}
\caption{Prior ranges for the pMSSM parameters, over which we perform our scans. All masses and trilinear couplings are given in\tev.\\ $^{\ast}$ In order to avoid generating a large number of points strongly disfavoured by the LHC 
we impose an additional cutoff on the physical gluino mass, $\mgluino\geq 0.75\tev$.}
\label{table:MSSMparams}
\end{center}

\end{table}

For scanning we use the package BayesFITS\cite{Fowlie:2012im,Kowalska:2012gs,Kowalska:2013hha,Fowlie:2013oua}, 
which interfaces several publicly available tools to 
direct the scanning procedure and calculate physical observables. 
The sampling is performed by \texttt{MultiNest}\cite{Feroz:2008xx} with 20,000 live points. 
The evidence tolerance is set to 0.0001 so that the stopping criterion is not reached before we collect a number of points 
deemed adequate for our purposes.
We use $\tt SoftSusy \,v.3.3.9$\cite{Allanach:2001kg} to calculate the mass spectrum. 
Higher-order corrections to the Higgs mass are calculated with $\tt FeynHiggs\ v.2.10.0$\cite{Heinemeyer:1998yj,Heinemeyer:1998np,Degrassi:2002fi,Frank:2006yh,Hahn:2013ria}.
$\tt FeynHiggs$ is interfaced with $\tt HiggsSignals$\cite{Bechtle:2013xfa} and $\tt HiggsBounds$\cite{Bechtle:2008jh,Bechtle:2011sb,Bechtle:2013wla} 
to evaluate the constraints on the Higgs sector. $\tt SuperISO\ v.3.3$\cite{Mahmoudi:2008tp} 
is used to calculate \brbxsgamma, \brbsmumu, \brbutaunu, and \deltagmtwomu.
$M_W$, \sinsqeff, \delmbs\ are calculated using $\tt FeynHiggs$. 
Dark matter observables \abundchi, \sigsip, \sigsdp, and \sigv, the annihilation branching ratios, 
the photon fluxes for CTA and Fermi-LAT, the neutrino-induced muon flux for IceCube,
and the positron flux are computed using $\tt micrOMEGAs\ v.3.5.5$\cite{Belanger:2013oya}. 

The scans are subject to a set of constraints, applied through a global likelihood function $\mathcal{L}$.
The list of constraints, central values, theoretical and experimental uncertainties are 
presented in Table~\ref{tab:exp_constraints}. We assume Gaussian distributions for the constraints, 
with the exception of those on the Higgs sector, which are imposed through $\tt HiggsSignals$ and $\tt HiggsBounds$, and the constraints
on \sigsip\ from LUX\cite{Akerib:2013tjd}. The LUX bound, which slightly improved on the limit from XENON100\cite{Aprile:2012nq}, 
is included in the likelihood function following the procedure described in detail in\cite{Cheung:2012xb,Fowlie:2013oua,Kowalska:2014hza,Roszkowski:2014wqa}. 
Additionally, we impose 95\%~C.L. lower bounds
from direct searches at LEP\cite{Agashe:2014kda}, smeared with 5\% theoretical errors. The limits are given in Eq.~(2) of Ref.\cite{Fowlie:2013oua}, 
with the exception of the limit on the neutralino mass 
that has been replaced here by the LEP limit on the invisible $Z$ width, $\Gamma(Z\rightarrow\chi\chi)$\cite{Agashe:2014kda}.
The points are gathered through several scans with both linear and log priors for the mass parameters.

\begin{table}[t]
\begin{center}
\begin{tabular}{|c|c|c|c|c|}
\hline
Constraint & Mean & Exp. Error & Th. Error & Ref. \\
\hline
Higgs sector & See text. & See text. & See text. & \cite{Bechtle:2013xfa,Bechtle:2008jh,Bechtle:2011sb,Bechtle:2013wla} \\
\hline
LUX & See\cite{Kowalska:2014hza,Roszkowski:2014wqa}. & See\cite{Kowalska:2014hza,Roszkowski:2014wqa}. & See\cite{Kowalska:2014hza,Roszkowski:2014wqa}. & \cite{Akerib:2013tjd}\\
\hline
\abunchi\ & 0.1199 & 0.0027 & 10\% & \cite{Ade:2013zuv}\\
\hline
\sinsqeff\ & 0.23155 & 0.00015 & 0.00015 &  \cite{Beringer:1900zz} \\
\hline
$\brbxsgamma\times 10^4$ & 3.43 & 0.22 & 0.21 & \cite{bsgamma} \\
\hline
$\brbutaunu \times 10^4$ & 0.72 & 0.27 & 0.38 & \cite{Adachi:2012mm} \\
\hline
\delmbs\ & 17.719~ps$^{-1}$ & 0.043~ps$^{-1}$ & 2.400~ps$^{-1}$ & \cite{Beringer:1900zz} \\
\hline
$M_W$ & $80.385\gev$ & $0.015\gev$ & $0.015\gev$ & \cite{Beringer:1900zz} \\
\hline
$\brbsmumu \times 10^9$ & 2.9 & $0.7$ & 10\% & \cite{Aaij:2013aka,Chatrchyan:2013bka} \\
\hline
$\Gamma(Z\rightarrow\chi\chi)$ & $\leq 1.7\mev$ & $0.3$ & -- & \cite{Agashe:2014kda} \\
\hline
\end{tabular}
\caption{The experimental constraints applied in this study.}
\label{tab:exp_constraints}
\end{center}
\end{table}%

We do not directly impose bounds on sparticle masses from direct SUSY searches at the LHC. 
As was explained in \refsec{sec:intro} we are particularly interested here on neutralino masses predominantly in the range
of a few hundred\gev\ to a few\tev, which are outside the reach of the LHC. 
The implementation of LHC searches in the likelihood function is for the pMSSM a 
numerically intensive task\cite{Fowlie:2013oua,Kowalska:2013ica} that goes beyond our purpose here.
We will explicitly mention in the text any situation in which a potential conflict with the limits from the LHC arises.    

Although some of the scans are also driven by the constraint on the anomalous magnetic moment of the muon, \deltagmtwomu\cite{Bennett:2006fi,Miller:2007kk},
we present our results irrespective of whether the measurement of \deltagmtwomu\ is satisfied, as the result is 
still somewhat controversial and favours the low-mass neutralino region, which is starting to show some tension with 
data from the LHC.  

When appropriate, we present results for two cases. 
In the first, $\abundchi\simeq 0.12$, we consider a Gaussian distribution for the relic density of DM, 
with experimental and theoretical uncertainties added in quadrature. 
In the second case, $\abundchi\lesssim 0.12$, the relic density is imposed as an upper bound only, by means of a half Gaussian likelihood. 

\section{\label{sec:pointchar}Neutralino properties and benchmark points}

In this and the following sections we show in the plots only the points
the satisfy the constraints of Table~\ref{tab:exp_constraints} at the 95\%~C.L., i.e., 
we select $\Delchisq\leq 5.99$ from the current best-fit point,
where $\chisq=-2\ln(\mathcal{L}/\mathcal{L}_{\textrm{max}})$\,.

\begin{figure}[t]
\centering
\subfloat[]{
\label{fig:a}
\includegraphics[width=0.47\textwidth]{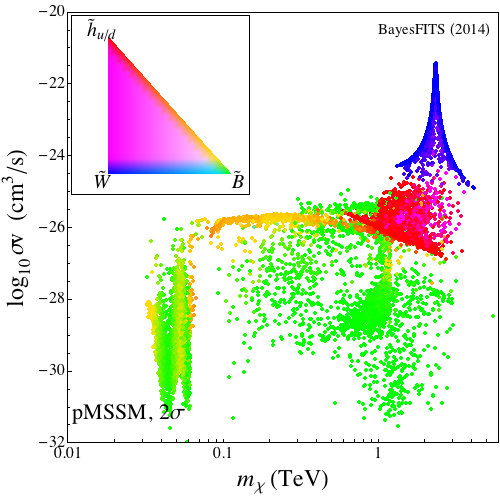} 
}
\hspace{0.01\textwidth}
\subfloat[]{
\label{fig:b}
\includegraphics[width=0.47\textwidth]{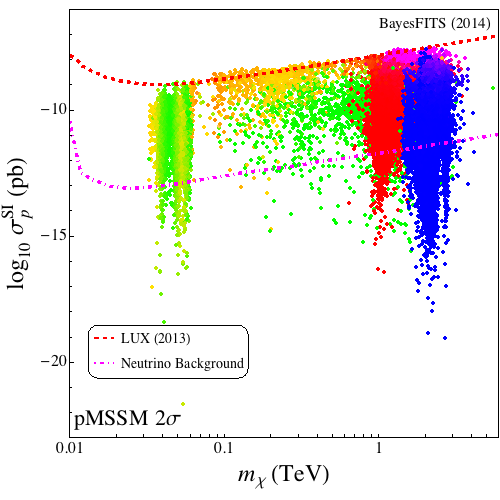}
}
\caption{\footnotesize\protect\subref{fig:a} The distribution of the points with $\Delchisq\leq 5.99$ (see Table~\ref{tab:exp_constraints}) in the (\mchi, \sigv) plane.
The colour coding identifies the composition of the lightest neutralino. Pure states
are shown in green for the bino~($\tilde{B}$), blue for the wino~($\tilde{W}$), 
and red for the higgsino~($\tilde{h}_{u/d}$). Admixtures are shown with intermediate colours in accordance with the legend. 
\protect\subref{fig:b} Same as \protect\subref{fig:a} but in the (\mchi, \sigsip) plane. 
The dashed red line shows the 90\%~C.L. bound from LUX\cite{Akerib:2013tjd}, included in the likelihood. 
The dot-dashed magenta line shows the onset of the irreducible atmospheric and diffuse supernova neutrino background\cite{Cabrera:1984rr,Monroe:2007xp,Billard:2013qya}.}
\label{fig:composition}
\end{figure}

In \reffig{fig:composition}\subref{fig:a} we show the distribution of our scan points in the  
(\mchi, \sigv) plane for the case where the LSP saturates the relic abundance, $\abundchi\simeq 0.12$\,.
We remind the reader that $\sigv=\langle\sigv\rangle|_{p\rightarrow 0}$\,.
The colour code gives the composition of the lightest neutralino.
The equivalent distribution in the (\mchi, \sigsip) plane is shown in \reffig{fig:composition}\subref{fig:b}.
The LUX bound on \sigsip\ is included in the likelihood; as a consequence in \reffig{fig:composition}\subref{fig:b} almost no points lie above the 
90\%~C.L. limit, shown here with a dashed red line for clarity. 

As is well known, the neutralino mass and composition
are determined by the relic density because it is a strong constraint with a relatively small uncertainty. 
The points of the elongated, almost vertical branches at $\mchi<100\gev$ 
belong to the $Z$- and $h$-resonance ``regions"\cite{Ellis:1989pg}. 
The neutralino mass is approximately half the mass of the $Z$ boson or of the lightest Higgs, 
so that resonant annihilation in the early Universe leads to the correct relic density. 
The neutralino is predominantly bino-like with a small admixture of higgsino that does not exceed $\sim 40\%$.
Because of their relatively low mass, neutralinos in this region will be the first among the SUSY particles to be tested 
at the LHC 14\tev\ run, possibly even through direct DM production as in the monojet/monophoton searches, 
which provide limits that do not depend on the presence of light charginos or sleptons in the spectrum. On the other hand, because of their 
suppressed present-day annihilation cross section these points are in principle not very interesting 
for indirect detection.

As the neutralino mass increases, $\mchi>100\gev$, predominantly bino-like LSPs (in green) satisfy the relic density
through different well-understood mechanisms. From the left to the right, the point models are characterised by ``bulk-like" annihilation to leptons\cite{Griest:1989zh,Drees:1992am}, slepton/neutralino co-annihilation\cite{Ellis:1998kh}, or resonance with heavy $A/H$ Higgs 
bosons\cite{Drees:1992am}.\footnote{A more detailed description of the mass ranges associated with each mechanism 
in a 9-dimensional low-scale parametrisation of the MSSM (p9MSSM) can be found in\cite{Fowlie:2013oua}.}

In the range $0.1\tev\lesssim\mchi\lesssim 1\tev$, neutralinos characterised by a mixed bino/higgsino composition   
satisfy the relic abundance partially through annihilation to gauge bosons, but also through off-shell $Z$ exchange
into top quarks, which can become dominant above the threshold since the other fermionic final states are suppressed by helicity conservation. 
The WIMPs in this region generally feature a large spin-independent cross section, so that if they have not been excluded by XENON100/LUX 
they are situated just below the 90\%~C.L. bound in \reffig{fig:composition}\subref{fig:b}.
Some of the points surviving the bound show significant higgsino levels but 
a relative sign difference between $\mu$ and $M_1$ suppresses the coupling of the neutralino to the lightest Higgs boson. 
Additionally, cancellations between the 
heavy and light Higgs diagrams reduce \sigsip\ giving rise to known ``blind spots"\cite{Ellis:2000ds,Ellis:2000jd,Ellis:2005mb,Baer:2006te,Huang:2014xua}. 
The position and depth of these spots depend on the relative sign of $\mu$ and $M_1$ and generally require that the masses of the heavy Higgs bosons are not 
much above the TeV scale.

Note also that in the pMSSM there exists the additional possibility of augmenting neutralino annihilations in the early Universe 
with co-annihilations with a squark or gluino.
This allows combinations of higgsino and bino components that can effectively evade the direct detection experiments.
One can see some examples of this in the mixed points at $\mchi\gsim1\tev$.
For indirect detection this scenario can feature a relatively large annihilation cross section but if co-annihilations 
are the dominant mechanism then this can be reduced below the projected sensitivity of CTA.
  
For masses equal and slightly larger than 1\tev, almost pure higgsinos naturally satisfy the relic density and direct detection 
constraints simultaneously. 
Figure~\ref{fig:composition}\subref{fig:a} shows that for these points (in red) the present-day annihilation 
cross section is large enough to be of interest to CTA. Pure higgsinos can lead to the correct relic density 
by quite a diversified set of annihilation mechanisms in the early Universe. Annihilation to gauge/Higgs bosons and chargino co-annihilation 
are the most natural options for masses on and above 1\tev, but one can see in \reffig{fig:composition}\subref{fig:a} 
some other higgsino points at 2--2.5\tev, for which the above-mentioned coannihilations with stops or gluinos effectively decrease the relic abundance. 
Conversely, for lower mass higgsinos, with \mchi\ as small as $\sim 600\gev$, 
mass degeneracies with sleptons and squarks increase the number of degrees of freedom at freeze-out, 
thus effectively boosting the value of the relic density with respect to the case without degeneracy\cite{Edsjo:1997bg}.
Additionally, one can see a large set of higgsino points above the ``canonical" thermal relic value for the cross section. 
There, \sigv\ is enhanced by a large resonant effect with the heavy Higgs bosons $A/H$ that was thermally washed out in the early Universe (for an explanation
see, e.g., Appendix~B of Ref.\cite{Fowlie:2013oua}).
  
An almost pure wino LSP (in blue) can naturally satisfy the relic density for masses around 2.8--3\tev\cite{Hryczuk:2010zi}.
As was the case for the nearly pure higgsino, winos can have masses that extend over a much larger interval,
$1.6\tev\lesssim\mchi\lesssim 4\tev$ thanks to coannihilations with squarks 
and gluinos, which reduce the relic density, or mass degeneracies with lighter scalar particles, which can increase it.

One can see in \reffig{fig:composition}\subref{fig:a} that Early Universe annihilation and co-annihilation 
of the winos receive a boost from the Sommerfeld enhancement\cite{Hisano:2006nn,Cirelli:2007xd,Hryczuk:2010zi}. 
We incorporated this effect as a correction factor to \abundchi, which we extrapolated 
from Fig.~5 of Ref.\cite{Hryczuk:2010zi} and from Ref.\cite{Cirelli:2007xd}.  
We also modified the present-day annihilation cross section
by the rescaling factors derived for different final states in\cite{Hryczuk:2011vi} for the case of wino DM.
We smeared the rescaling factors with a quadratic function of $\Delta m_{\chi}=m_{\charone}-\mchi$
to extend the Sommerfeld effect to the few mixed wino/higgsino points featured in our scan.
The effects of the Sommerfeld enhancement can be seen in the peculiar ``Eiffel Tower" shaped 
resonance in \reffig{fig:composition}\subref{fig:a} for wino (in blue) and mixed wino/higgsino LSPs (in magenta).

Additionally, the spectrum also features an enhanced $\chi \chi \rightarrow \gamma \gamma$ cross section.
Because of this, scenarios dominated by wino DM are in conflict with data from the current generation of $\gamma$-ray 
telescopes, namely they are excluded\cite{Cohen:2013ama,Fan:2013faa,Hryczuk:2014hpa} at the 95\%~C.L. by the H.E.S.S. search for $\gamma$-ray lines\cite{Abramowski:2013ax},
with the exception of cases where the halo profile is very flat. We will show in \refsec{sec:current} that the same result applies to the wino-like 
points in our scans.

\begin{figure}[t]
\centering
\includegraphics[width=0.50\textwidth]{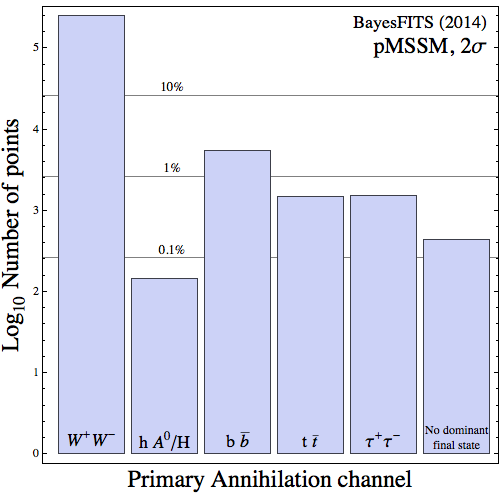}
\caption{\footnotesize Bar chart of the points in the scan ($\Delchisq\leq 5.99$) according to their 
primary annihilation channel (the primary final state is the one with the largest branching fraction).}
\label{fig:histo}
\end{figure}

The bar chart in \reffig{fig:histo} shows the number of points with $\Delchisq\leq 5.99$ organised by their predominant annihilation channel. 
We classify each point in terms of which final state has the largest branching fraction but for the majority of points 
annihilation will occur to several different final states. 
One can see that the largest number predominantly annihilates into $W^+W^-$, 
which is the preferred final state for both the higgsino and wino high-density regions in \reffig{fig:composition}. 

\begin{figure}[t]
\centering
\subfloat[]{%
\label{fig:a}%
\includegraphics[width=0.45\textwidth]{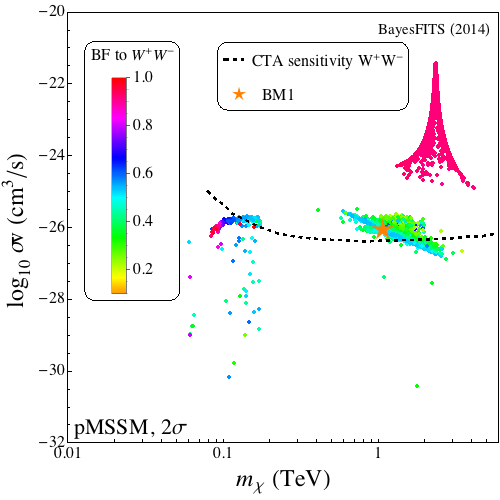}
}%
\hspace{0.07\textwidth}
\subfloat[]{%
\label{fig:b}%
\includegraphics[width=0.45\textwidth]{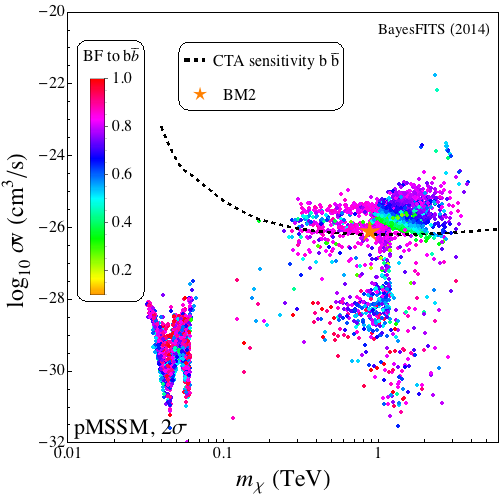}
}%

\subfloat[]{%
\label{fig:c}%
\includegraphics[width=0.45\textwidth]{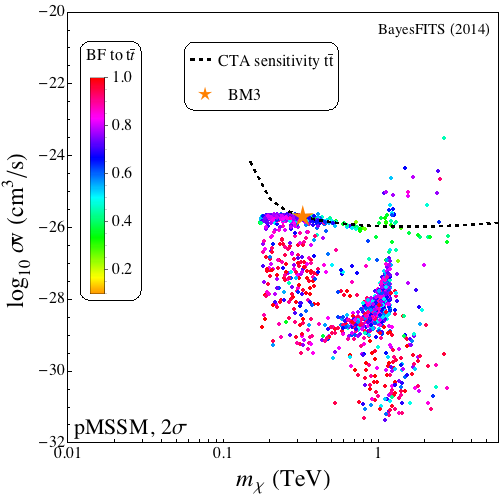}
}%
\hspace{0.07\textwidth}
\subfloat[]{%
\label{fig:d}%
\includegraphics[width=0.45\textwidth]{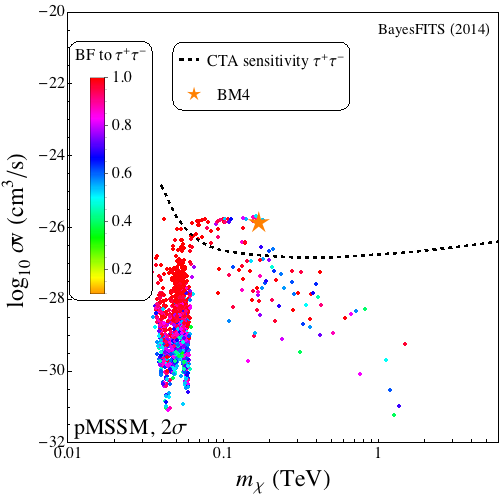}
}%
\caption{\footnotesize\protect\subref{fig:a} Distribution in the (\mchi, \sigv) plane of the points with $\Delchisq\leq 5.99$ (see Table~\ref{tab:exp_constraints}) 
annihilating primarily into $W^+W^-$. 
The colour code shows the value of the $W^+W^-$ branching fraction for each point. 
The dashed black line shows the projected sensitivity of CTA to points with $\textrm{BR}(\chi\chi\rightarrow W^+W^-)=100\%$, which we derive 
in Appendix~\ref{sec:app}. The orange star marks the position of BM1. \protect\subref{fig:b} 
Same as \protect\subref{fig:a} but for $b\bar{b}$\,. The orange star marks the position of BM2. \protect\subref{fig:c} 
Same as \protect\subref{fig:a} but for $t\bar{t}$\,. The orange star marks the position of BM3.
\protect\subref{fig:d} Same as \protect\subref{fig:a} but for $\tau^+\tau^-$\,. The orange star marks the position of BM4.}  
\label{fig:finalstate}
\end{figure}

In \reffig{fig:finalstate}\subref{fig:a} we show the distribution in the (\mchi, \sigv) plane of the points
in the first column of \reffig{fig:histo} (largest branching ratio to $W^+W^-$).
The colour code shows the actual value of the branching ratio for each point.
The distributions of the points in the 3rd ($b\bar{b}$), 4th ($t\bar{t}$), and 5th ($\tau^+\tau^-$) columns of \reffig{fig:histo}
are shown in Figs.~\ref{fig:finalstate}\subref{fig:b}, \ref{fig:finalstate}\subref{fig:c}, and \ref{fig:finalstate}\subref{fig:d},
respectively. To allow a direct comparison, in the panels of \reffig{fig:finalstate} we indicate with a dashed black line 
the expected reach of CTA in the relative final state. We will discuss in detail in \refsec{sec:CTA} and Appendix~\ref{sec:app} the sensitivity of CTA.  
In \reffig{fig:finalstate} we also mark with orange stars the positions of a few benchmark points for indirect detection, which we present in Table~\ref{tab:benchmarks} 
and whose characteristics we describe below.

\begin{table}[t]
\begin{center}
\begin{tabular}{|c|c|c|c|c|}
\hline
Benchmark & BM1 & BM2 & BM3 & BM4 \\
\hline
$M_1$ & Decoupled & 878 & 353 & -189 \\
\hline
$M_2$ & Decoupled & Decoupled & 1500 & Decoupled \\
\hline
$\mu$ & -1057 & 1300 & -369 & 218 \\
\hline
\ma\ & Decoupled & 1832 & 2500 & 754 \\
\hline
\tanb\ & 58.0 & 49.4 & 26.6 & 30.0 \\
\hline
$m_{\stau_R}$ & Decoupled & Decoupled & Decoupled & 253 \\
\hline
\mchi\ & 1077 & 883 & 328 & 172 \\
\hline
Composition & $|N_{\tilde{B}}|^2=0.000015$ & $|N_{\tilde{B}}|^2=0.996$ & $|N_{\tilde{B}}|^2=0.730$ & $|N_{\tilde{B}}|^2=0.777$ \\
 & $|N_{\tilde{W}}|^2=0.000076$ & $|N_{\tilde{W}}|^2=4 \times 10^{-7}$ & $|N_{\tilde{W}}|^2=0.00043$ & $|N_{\tilde{W}}|^2=0.00084$ \\ 
 & $|N_{\tilde{h}_u}|^2=0.5$ & $|N_{\tilde{h}_u}|^2=0.00313$ & $|N_{\tilde{h}_u}|^2=0.156$ & $|N_{\tilde{h}_u}|^2=0.142$ \\ 
 & $|N_{\tilde{h}_d}|^2=0.5$ & $|N_{\tilde{h}_d}|^2=0.00136$ & $|N_{\tilde{h}_d}|^2=0.114$ & $|N_{\tilde{h}_d}|^2=0.080$ \\  
\hline
\abundchi\ & 0.119 & 0.121 & 0.119 & 0.122 \\
\hline
Main & \footnotesize{chargino co-ann.} & \footnotesize{$A$-resonance} & \footnotesize{$s$ ch. off-shell $Z$} & \footnotesize{$t$-ch. exch. \stau,\charone,\neuttwo} \\
mechanism & $\chi\chi^\pm \rightarrow$ SM & $\chi\chi\rightarrow b\bar{b}$ & $\chi\chi\rightarrow t\bar{t}$ &  $\chi\chi\rightarrow\tau^+\tau^-,$ \\
\abundchi\ & & & &  $W^+W^-, t\bar{t}, ZZ$ \\
\hline
\sigv\ ($\textrm{cm}^3$/s) & $8.45 \times 10^{-27} $&$7.59 \times 10^{-27} $ &$1.96 \times 10^{-26} $  & $1.34 \times 10^{-26} $  \\
\hline
Branching & 53.8\% $W^+W^-$ & 79.7\% $b\bar{b}$ & 74.1\% $t\bar{t}$ & 36.4\% $\tau^+\tau^-$ \\
fractions & 45.1\% $ZZ$ & 20.2\% $\tau^+\tau^-$ & 12.9\% $W^+W^-$ & 31.1\% $W^+W^-$ \\
 & 0.983\% $t\bar{t}$ & 0.1\% $d\bar{d}/s\bar{s}$ & 10.4\% $ZZ$ & 24.5\% $ZZ$ \\ 
 & 0.116\% $Zh$ &  & 1.9\% $Zh$ & 7.2\% $b\bar{b}$ \\  
\hline
$\sigma_{\gamma\gamma} v$ ($\textrm{cm}^3$/s) & $5.31 \times 10^{-29}$  & $1.04 \times 10^{-33}$ & $2.32 \times 10^{-30}$  & $1.58 \times 10^{-30}$\\
\hline
$\sigma_{Z\gamma} v$ ($\textrm{cm}^3$/s) & $4.75 \times 10^{-29}$ &  $3.04 \times 10^{-34}$ & $1.86 \times 10^{-29}$ & $1.11\times 10^{-29}$\\
\hline
\sigsip\ (pb) & $4.73 \times 10^{-12}$  & $1.65 \times 10^{-10}$ & $3.37 \times 10^{-9}$ & $1.14\times10^{-9}$\\
\hline
\sigsdp\ (pb) & $7.75 \times 10^{-9}$ & $ 1.17 \times10^{-7}$ & $7.51 \times 10^{-5}$ & $1.55\times 10^{-4}$\\
\hline
\end{tabular}
\caption{Benchmark points for indirect detection. SUSY masses are given in\gev. ``Decoupled" stands for any mass value above 5000\gev.}
\label{tab:benchmarks}
\end{center}
\end{table}%

In \reffig{fig:finalstate}\subref{fig:a} one can see that the $\sim1\tev$ higgsino region 
must feature very substantial branching fractions to subdominant final states other than $W^+W^-$. 
In general, higgsinos can often have an almost equal branching fraction 
to $ZZ$, $Zh$ or $Z A$, through $t$-channel exchange of the neutral, almost degenerate second-lightest neutralino.
In \reffig{fig:finalstate}\subref{fig:b} one can see that points annihilating into $b\bar{b}$ final states
are very common and spread over the entire range of the parameter space.
These are typical of the $A$-resonance mechanism, when $\ma\approx2\mchi$, and they are characterised by moderate-to-large \tanb.
Note that, for $\tanb\lesssim 8$, points characterised by resonance with the heavy Higgses 
prefer to annihilate to $t\bar{t}$ instead. One can see these points in \reffig{fig:finalstate}\subref{fig:c}
for masses  $\mchi\gsim 500\gev$ (in the same part of the figure many points are also characterised by stop co-annihilation in the early Universe). 
The points at $\mchi< 500\gev$ in \reffig{fig:finalstate}\subref{fig:c} are instead points that satisfy the relic density 
through selectron (or smuon) co-annihilation, or annihilation to tops through an off-shell $Z$, 
for which $s$-wave production of light fermions is helicity-suppressed.   

In \reffig{fig:finalstate}\subref{fig:d} one can see that many of the points that satisfy the relic density at the $Z/h$-resonance,
$\mchi\simeq 45-65\gev$, very predominantly annihilate in the present-day Universe to $\tau^+\tau^-$ and their cross section is depleted
with respect to the ``canonical" thermal value.
This is because, on the one hand the Boltzman distribution for WIMPs in this range is sharply peaked at momentum $p>0$
so that one expects a drop in the cross section now, when $p\rightarrow 0$;
on the other hand, the lightest stau happens to be for these points light enough ($m_{\stau_1}\simeq100-500\gev$) to allow significant annihilation 
to taus through stau exchange in the $t$ channel.
Note that even if the points in this region are predominantly bino-like they need a non-negligible amount of mixing with the higgsino to 
efficiently annihilate in the early Universe. Thus, for many of these points the lightest chargino, \charone, and the 
second-lightest neutralino, \neuttwo\,, have masses not much far above \mchi, and they are therefore 
very likely to be already excluded by 3-lepton searches at the LHC\cite{Aad:2014vma,Aad:2014nua,Khachatryan:2014qwa,Khachatryan:2014mma}. 
However, there are several other points in this region for which $m_{\charone}\gsim 450\gev$ 
so that they are not currently excluded and remain viable scenarios. 
The 14\tev\ run at the LHC should definitively probe these points. 

We present four benchmark points for DM detection in Table~\ref{tab:benchmarks}, 
each one is marked by an orange star in the panels of \reffig{fig:finalstate}.
They are chosen such that they are not presently excluded but represent scenarios within the projected reach of CTA.

Benchmark point~1 (BM1) belongs to the $\sim1\tev$ higgsino region. 
The dominant final states are $W^+W^-$ and $Z Z$. The $\gamma\gamma$ and $Z\gamma$ cross sections are 
suppressed compared to the continuous spectrum but remain relatively large due to the higgsino nature of the neutralino 
and the associated light chargino that appears in the loops. 
Benchmark point~2 (BM2) is a typical $A$-resonance point annihilating predominately to $b\bar{b}$. 
It is the only benchmark point without sizeable higgsino-bino mixing. Due to this the $\gamma\gamma$ and $Z\gamma$ 
cross sections are heavily suppressed.
Benchmark point~3 (BM3) is a mixed bino-higgsino neutralino. The $t\bar{t}$ final state is produced 
non-resonantly and thus requires large mixing to increase the coupling of the neutralino to the $Z$ boson. 
The chargino associated with the higgsino component is relatively light and induces the sub-dominant annihilation channels to $W^+W^-$ and $ZZ$. 
Finally, benchmark point~4 (BM4) is another mixed bino-higgsino neturalino similar to BM3 although its 
dominant annihilation channel is now mediated by a light stau leading to a $\tau^+\tau^-$ final state.
For BM4 the neutralinos in the early Universe also annihilate into top quarks at threshold, 
but top production is suppressed for $p\rightarrow 0$ because the neutralino is slightly too light.

It should be noted that both mixed bino-higgsino scenarios, BM3 and BM4, 
have a spin-independent scattering cross section close to the current limit from LUX, thus 
making these scenarios highly visible in future direct detection experiments.
BM2 may be detectable at future 1-tonne direct detection experiments, while BM1 lies in a direct detection blind spot.

\section{Prospects for dark matter detection\label{sec:results}}

We present here the prospects for DM detection in the pMSSM. 
In \refsec{sec:current} we summarise the current status of indirect-detection bounds on the parameter space
of the pMSSM. We show the case where the neutralino saturates the relic abundance, $\abundchi\simeq 0.12$, and the case $\abundchi\lesssim 0.12$. 
In \refsec{sec:CTA} we present our estimate of the sensitivity reach of CTA. 
We give results for the Einasto and NFW DM halo profiles, obtained through the procedure 
described in detail in Appendix~\ref{sec:app}. Finally, 
in \refsec{sec:comp} we compare the reach of CTA with the sensitivity of other direct and indirect detection experiments.

\subsection{Current indirect detection bounds on the pMSSM\label{sec:current}}

It was shown, e.g., in Ref.\cite{Fowlie:2013oua} for the p9MSSM and in Ref.\cite{Cahill-Rowley:2014boa} for the p19MSSM that
the bounds from the continuous $\gamma$-ray spectrum
from dSphs at Fermi-LAT\cite{Ackermann:2013yva} are too weak to affect the MSSM parameter space when the PLANCK value\cite{Ade:2013zuv} on 
\abundchi\ is imposed.
The limits can however become important in scenarios where the relic abundance 
is satisfied as an upper bound only. 
This opens up large regions of the parameter space, characterised by lighter higgsino and wino LSPs that annihilate away 
effectively in the early Universe and have a much suppressed relic abundance. 
These solutions could correspond to a scenario with two-component 
DM, in which case the sensitivity of Fermi-LAT should be rescaled according to the square of the density, $R^2$, to account 
for the reduced present-day density of neutralinos ($R=\Omega_{\chi}/\Omega_{\textrm{Planck}}$).  
Alternatively, one can assume that the neutralino represents the entirety of the DM 
and the correct abundance is fixed by invoking some additional mechanism, e.g., freeze-in\cite{Hall:2009bx}. 
In this case one obtains much stronger limits from indirect detection than in the previous scenario because 
rescaling by $R^2$ is not necessary\cite{Williams:2012pz}.
Only in this case does the limit from dSphs have significant impact on the parameter space.
However, the limit from the $\gamma$-ray line search at H.E.S.S.\cite{Abramowski:2013ax} can also affect a small part of the parameter space 
when one applies rescaling.

\begin{figure}[t]
\centering
\subfloat[]{%
\label{fig:a}%
\includegraphics[width=0.45\textwidth]{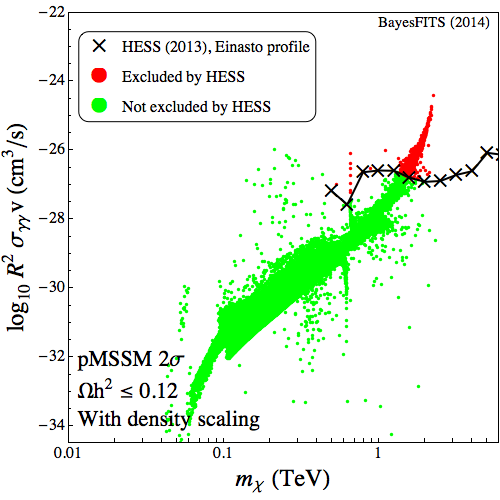}
}%
\hspace{0.07\textwidth}
\subfloat[]{%
\label{fig:b}%
\includegraphics[width=0.45\textwidth]{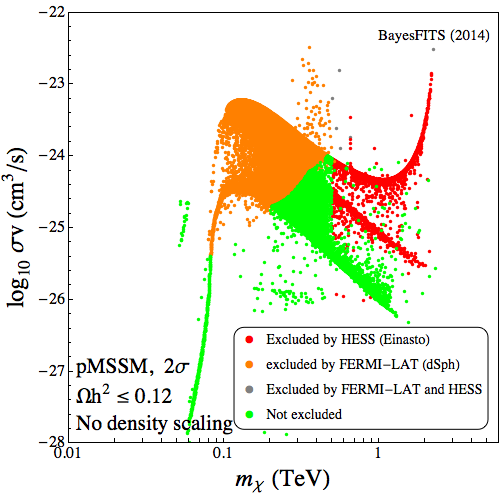}
}%
\caption{\footnotesize Current status of indirect detection bounds on the pMSSM for $\abundchi\lesssim0.12$. 
\protect\subref{fig:a} The points excluded at the 95\%~C.L. by $\gamma$-ray line searches at H.E.S.S.\cite{Abramowski:2013ax} 
(in red) in the (\mchi, $R^2\cdot\sigma_{\gamma\gamma}v$) plane.
 \protect\subref{fig:b} The points excluded at the 95\%~C.L. by Fermi-LAT dSphs\cite{Ackermann:2013yva} (orange) 
 and H.E.S.S. $\gamma$-ray line search (red) in the (\mchi, \sigv) plane (case without rescaling).}  
\label{fig:norel}
\end{figure}

In \reffig{fig:norel}\subref{fig:a} we show in red the points excluded at the 95\%~C.L. by 
the H.E.S.S. search under the assumption of rescaling by $R^2$.
In \reffig{fig:norel}\subref{fig:b} we show the points excluded at the 95\%~C.L. by Fermi-LAT dSphs (in orange) and the above mentioned 
H.E.S.S. $\gamma$-ray line search (in red) projected to the (\mchi, \sigv) plane in the case of no rescaling.
In the absence of additional guidance from theory, the real limit is likely to lie somewhere in between the exclusions of 
Figs.~\ref{fig:norel}\subref{fig:a} and \ref{fig:norel}\subref{fig:b}.

Going back to the case where the neutralino saturates the relic density, 
the strongest indirect limits on the spin-dependent scattering cross section
come from IceCube/DeepCore\cite{IceCube:2011aj,Aartsen:2012kia} and ANTARES\cite{Adrian-Martinez:2013ayv}, from observation of neutrinos from the Sun. 
It is well known\cite{Jungman:1995df} that \sigsdp\ can easily be measured through the relation between the DM solar capture rate and the annihilation rate, which should give rise to a high-energy neutrino spectrum and flux.

\begin{figure}[t]
\centering
\subfloat[]{%
\label{fig:a}%
\includegraphics[width=0.45\textwidth]{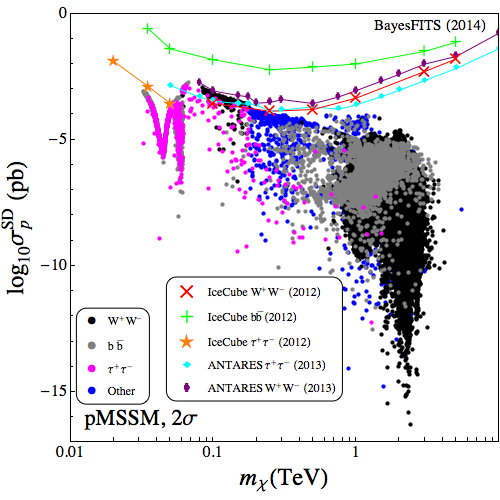}
}%
\hspace{0.07\textwidth}
\subfloat[]{%
\label{fig:b}%
\includegraphics[width=0.45\textwidth]{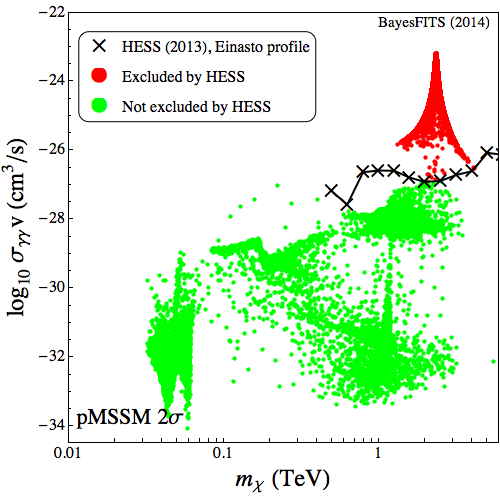}
}%
\caption{\footnotesize\protect\subref{fig:a} The current 90\%~C.L. limits on \sigsdp\ from IceCube\cite{IceCube:2011aj,Aartsen:2012kia} 
and ANTARES\cite{Adrian-Martinez:2013ayv}. 
The limits are presented for the $W^+W^-$, $b\bar{b}$, and $\tau^+\tau^-$ final states. Points are shown with $\Delchisq\leq 5.99$ (see Table~\ref{tab:exp_constraints}).
The colour code shows the primary annihilation final state 
for each point. \protect\subref{fig:b} The current 95\%~C.L. limits on $\sigma_{\gamma\gamma} v$ from H.E.S.S.\cite{Abramowski:2013ax}. 
In red we show the points that are excluded by the experimental bound, indicated with a solid black line.
Points are shown with $\Delchisq\leq 5.99$ (see Table~\ref{tab:exp_constraints}).}
\label{fig:current}
\end{figure}

In \reffig{fig:current}\subref{fig:a} we show the current limits 
from IceCube and ANTARES, provided for 100\% 
branching ratios to $W^+W^-$, $b\bar{b}$, and $\tau^+\tau^-$,
compared to the points of the pMSSM. The colour code shows the dominant
annihilation final state for each point, in the sense of \reffig{fig:histo}.

As was mentioned in \refsec{sec:pointchar} the $\gamma$-ray line search at H.E.S.S.\cite{Abramowski:2013ax}
is very effective in placing a strong bound on wino DM when $\abundchi\simeq 0.12$. We show
in \reffig{fig:current}\subref{fig:b} the current limit on the 
cross section under the assumption of the Einasto halo profile, 
compared to the pMSSM points. 
As was previously shown in\cite{Cohen:2013ama,Fan:2013faa,Hryczuk:2014hpa}, 
when one includes the Sommerfeld enhancement H.E.S.S. excludes at the 95\%~C.L. pure or almost pure wino DM,
with the exception of cases where a flat DM profile is assumed. 
(We repeat that we include the Sommerfeld enhancement as a smeared correction factor derived from Ref.\cite{Hryczuk:2011vi}.) 
Note that it has been recently shown\cite{Bauer:2014ula} that the estimate of the wino annihilation cross section is subject to 
a significant perturbative uncertainty. However, even when accounting for these uncertainties the H.E.S.S. 
limit is still quite constraining for this scenario.     
For the remainder of this paper we will not show in the plots the points excluded by H.E.S.S..

\begin{figure}[t]
\centering
\includegraphics[width=0.50\textwidth]{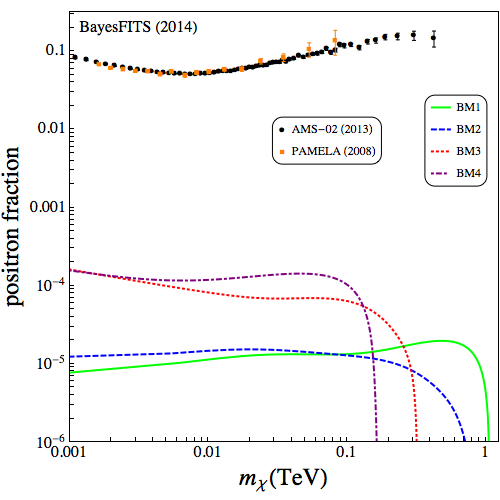}
\caption{\footnotesize The DM annihilation contribution to the positron fraction for the benchmark points of Table~\ref{tab:benchmarks} compared to the measured positron fraction at Pamela\cite{Adriani:2008zr} and AMS-02\cite{Aguilar:2013qda}.}
\label{fig:ams}
\end{figure}

In \reffig{fig:ams} we show the DM annihilation contribution to the positron fraction for the benchmark points of Table~\ref{tab:benchmarks}.
We compare our benchmark points to the measured positron fraction at Pamela\cite{Adriani:2008zr} and AMS-02\cite{Aguilar:2013qda}.
We use {\tt micrOMEGAs} to calculate the produced positron spectrum for each point and the positron flux at the Earth after propagation.
We use the default values in {\tt micrOMEGAs} for charged particle propagation.
To compare with the positron fraction we use the parametrisation of the primary electron background and secondary electron and positron fluxes following\cite{Baltz:1998xv,Moskalenko:1997gh}.
The fluxes in $\gev\ \text{cm}^{-2}\ \text{s}^{-1}\ \text{sr}^{-1}$ are given by,

\begin{equation}
\left(\frac{d\Phi}{dE}\right)_{\textrm{primary},\, e^-} = \frac{0.16 E^{-1.1}}{1 + 11 E^{0.9} + 3.2 E^{2.15}}\,,
\end{equation}
\begin{equation}
\left(\frac{d\Phi}{dE}\right)_{\textrm{secondary},\, e^-} = \frac{0.7 E^{0.7}}{1 + 110 E^{2.9} + 600 E^{2.9} + 580 E^{4.2}}\,,
\end{equation}
\begin{equation}
\left(\frac{d\Phi}{dE}\right)_{\textrm{secondary},\, e^+} = \frac{4.5 E^{0.7}}{1 + 650 E^{2.3} + 1500 E^{4.2} + 580 E^{4.2}}\,. 
\end{equation}

For all of the benchmarks the positron flux is many orders of magnitude too small to explain the anomalous positron fraction,
which is also true for the rest of the points.

\subsection{Sensitivity of CTA to the pMSSM\label{sec:CTA}}

The sensitivity of CTA is obtained by testing the background-only hypotesis with a likelihood function. 
The construction of the likelihood is detailed in Appendix~\ref{sec:app}. We use the most up to date\cite{JCarr} instrument response functions and background estimates provided by the CTA Collaboration\cite{montecarlo}. 
We assume an observation time of approximately 500 hours, and we provide results under the Einasto and NFW DM halo profiles, for which we calculate the $J$-factors.
 
\begin{figure}[t]
\centering
\subfloat[]{%
\label{fig:a}%
\includegraphics[width=0.45\textwidth]{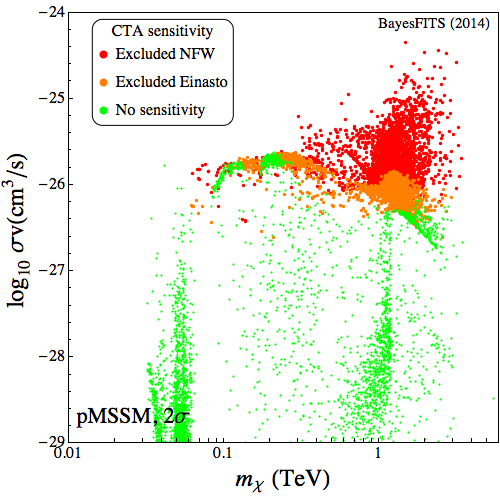}
}%
\hspace{0.07\textwidth}
\subfloat[]{%
\label{fig:b}%
\includegraphics[width=0.45\textwidth]{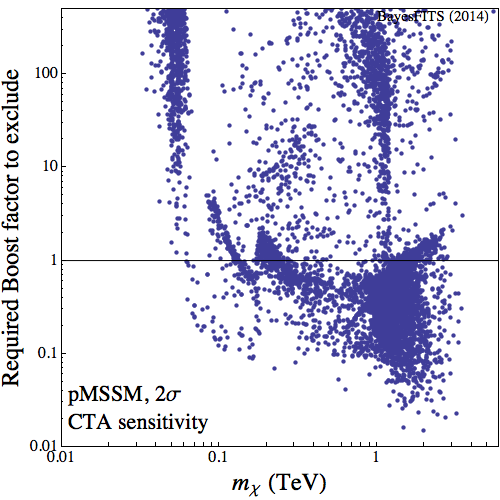}
}%
\caption{\footnotesize\protect\subref{fig:a} Sensitivity of CTA to the pMSSM in the (\mchi, \sigv) plane. 
Red points are within reach of CTA assuming an NFW profile, orange points assuming an Einasto profile, 
green points are beyond the sensitivity of CTA. Points are shown with $\Delchisq\leq 5.99$ (see Table~\ref{tab:exp_constraints}). 
The details of the calculation are given in Appendix~\ref{sec:app}.
\protect\subref{fig:b} The boost factor to \sigv, required for each point to be within the 95\%~C.L. sensitivity of CTA for the Einasto profile.}  
\label{fig:CTA1}
\end{figure}

We start with the case where the lightest neutralino makes up all of the DM, $\abundchi\simeq 0.12$. In~\reffig{fig:CTA1}\subref{fig:a} we show the 95\%~C.L. sensitivity of CTA in the (\mchi, \sigv) plane. 
We show points within the reach of CTA under the assumption of the NFW profile in red, 
while the points excluded under the Einasto profile but not the NFW are shown in orange. Green points lie beyond the reach of CTA for either profile. 
We repeat that we do not show in the plots the points excluded by H.E.S.S. (see \reffig{fig:current}\subref{fig:b}),
which feature a nearly pure wino LSP
or an admixture of wino and higgsino. 
Our improved analysis shows that the reach of CTA for SUSY is even more optimistic than our previous study had 
anticipated\cite{Roszkowski:2014wqa} and
under the Einasto profile CTA is bound to exclude at the $2\sigma$ level the $\sim 1\tev$ higgsino region of the pMSSM in almost its entirety,
effectively closing the window for heavy SUSY DM in many realistic models.

The point is emphasised in~\reffig{fig:CTA1}\subref{fig:b}, where we show the boost factor, $b_F$, required for each point to be within the 
95\%~C.L. sensitivity of CTA for the Einasto profile. Points below $b_F=1$ can be excluded without a boost factor. 
One can see that, in agreement with \reffig{fig:CTA1}\subref{fig:a}, the majority of points in the $\sim 1\tev$ higgsino region do not require any boost factor.

Incidentally, we want to underline here the power of the binned likelihood function especially in 
relation to the points that show a significant cross section to $\gamma\gamma$. 
This is what happens for the points in red and orange at $\mchi<300\gev$ in \reffig{fig:CTA1}\subref{fig:a}, 
which are projected to be excluded under one or the other profile assumption. 
Their $\gamma$-ray flux effectively gives rise to an excess in one or a few particular bins, 
large enough to kill the likelihood function irrespectively of all other bins.

It has been suggested in the literature\cite{Salucci:2007tm,Nesti:2013uwa} that a better fit for the DM profile 
in spiral galaxies, including possibly the Milky Way, might be provided by the Burkert rather than the Einasto or NFW profiles.
We have checked that the Burkert profile assumption has the effect of weakening the projected reach of CTA in the (\mchi, \sigv) plane 
by a factor of $\sim 50$ with respect to the Einasto profile. Under the Burkert profile assumption \reffig{fig:CTA1}
would have to be modified as follows: on the one hand the $\sim 1\tev$ higgsino region would be outside of the reach of CTA;
on the other hand, the vast majority of the points with wino DM, which are excluded by H.E.S.S. under the Einasto assumption, 
would then be allowed and directly in reach of CTA. 
  
\begin{figure}[t]
\centering
\subfloat[]{%
\label{fig:a}%
\includegraphics[width=0.45\textwidth]{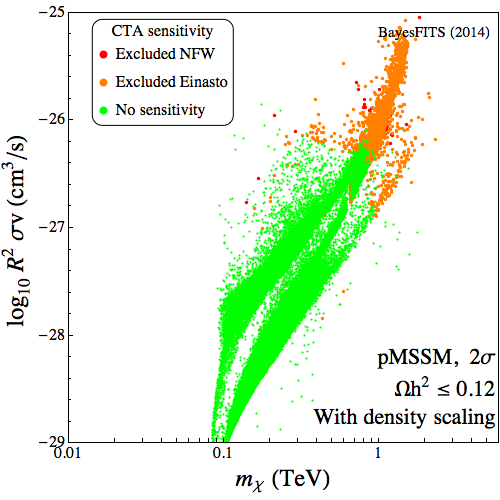}
}%
\hspace{0.07\textwidth}
\subfloat[]{%
\label{fig:b}%
\includegraphics[width=0.45\textwidth]{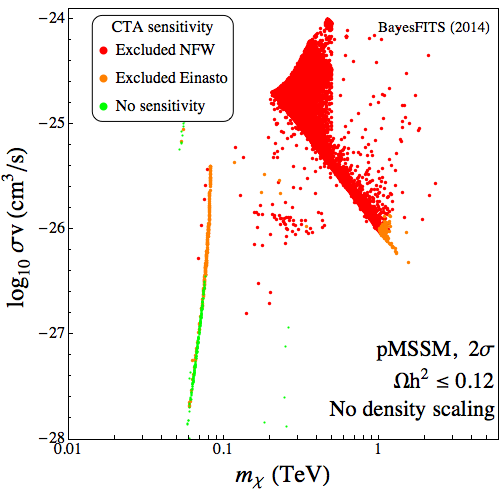}
}%
\caption{\footnotesize The reach of CTA in the pMSSM for the case $\abundchi\lesssim 0.12$. 
The colour code is the same as in \reffig{fig:CTA1}\protect\subref{fig:a}.
\protect\subref{fig:a} The 95\%~C.L. reach of CTA 
in the (\mchi, $R^2\cdot\sigv$) plane for the case with rescaling of the local density.  
\protect\subref{fig:b} The 95\%~C.L. reach of CTA in the (\mchi, \sigv) plane for the case without rescaling. 
}  
\label{fig:ctanorel}
\end{figure}

We now move on to describe the reach of CTA in the case where the neutralino does not saturate the relic density, $\abundchi\lesssim 0.12$.
Figure~\ref{fig:ctanorel}\subref{fig:a} shows the expected reach of CTA in the (\mchi, $R^2\cdot\sigv$) 
plane in the case of rescaling \sigv\ by the ratio (squared) of the neutralino density to the PLANCK value. 
We do not show in the plot the points excluded by Fermi-LAT dSphs and the $\gamma$-ray line search at H.E.S.S., see \reffig{fig:norel}.
One can see that CTA will probe regions of the parameter space beyond the limit presently set by H.E.S.S., especially 
if one assumes the Einasto halo profile. A few points (in red) are excluded by CTA also under assumption of the NFW profile 
but in that case much of the parameter space lies beyond the reach of CTA due to the rescaling factor $R^2$.

As was the case in \reffig{fig:norel}, when one assumes no rescaling the sensitivity of CTA can exclude a large fraction of the parameter space under both profile assumptions, as one can see  
in \reffig{fig:ctanorel}\subref{fig:b} where the CTA reach is shown in the (\mchi, \sigv) plane.

\subsection{Complementarity of CTA with other experiments\label{sec:comp}}

\begin{figure}[t]
\centering
\subfloat[]{%
\label{fig:a}%
\includegraphics[width=0.45\textwidth]{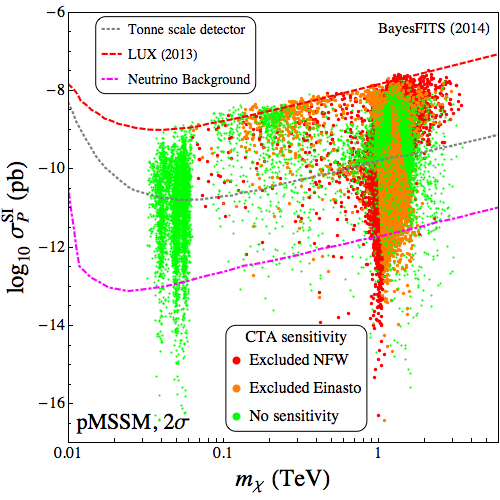}
}%
\hspace{0.07\textwidth}
\subfloat[]{%
\label{fig:b}%
\includegraphics[width=0.45\textwidth]{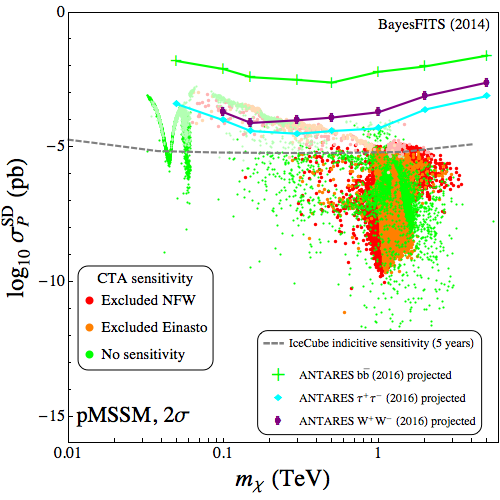}
}%
\caption{\footnotesize In all panels the colour code is the same as in~\reffig{fig:CTA1}\protect\subref{fig:a}. 
Points are shown with $\Delchisq\leq 5.99$ (see Table~\ref{tab:exp_constraints}).
\protect\subref{fig:a} The sensitivity of CTA to the pMSSM in the (\mchi, \sigsip) plane. 
The LUX 90\%~C.L. bound is shown a dashed red line. The projected sensitivity of 1-tonne detectors is shown as a dotted grey line. 
The onset of the atmospheric and diffuse supernova neutrino background is shown with a dot-dashed magenta line.
Note that points are plotted from red to green, showing a conservative estimate of the reach (least constrained points are always shown).
\protect\subref{fig:b} The sensitivity of CTA to the pMSSM in the (\mchi, \sigsdp) plane. 
Lighter shaded points are within the projected 5-year sensitivity of IceCube/DeepCore. The dashed grey line is indicative of IceCube's sensitivity.
The solid lines show the projected reach at ANTARES\cite{VBertin} for 2016 in the $b\bar{b}$ (green), $W^+W^-$ (purple), and $\tau^+\tau^-$ (cyan) final states.}  
\label{fig:comple1}
\end{figure}

In \reffig{fig:comple1}\subref{fig:a} we show the projected reach of CTA in the (\mchi, \sigsip) plane 
to compare it with direct detection experiments, which are sensitive to the spin-independent neutralino-proton cross section, \sigsip.
We use the same colour code as in \reffig{fig:CTA1}\subref{fig:a} for the sensitivity of CTA.
The projected sensitivity of \xenononet\cite{Aprile:2012zx}, indicative of the generic reach of 1-tonne 
detectors like, e.g., DEAP-3600\cite{Amaudruz:2014nsa} and LZ\cite{Malling:2011va}, is shown as a dashed grey line. 
The dot-dashed magenta line shows the onset of the irreducible background due to atmospheric and diffuse supernova neutrinos\cite{Cabrera:1984rr,Monroe:2007xp,Billard:2013qya}.
Since there is considerable overlap 
between the regions within the reach of CTA and those out of reach, 
it should be noted that the points are overlaid from the most constrained to the least constrained. 
Thus, the plots represent a conservative estimate of the reach of CTA in these planes. 

Figure~\ref{fig:comple1}\subref{fig:a} shows that over most of the pMSSM parameter space the reach of CTA is orthogonal to 
that of detectors directly measuring \sigsip. 
CTA will be able to probe the vast majority of the points that lie well beyond the reach of 1-tonne detectors and
even reach the region where the irreducible neutrino background will strongly curb sensitivity advances
for direct detection experiments.

In \reffig{fig:comple1}\subref{fig:b} we show the sensitivity of CTA compared 
to the projected 5-year sensitivity of the IceCube/DeepCore 86-string configuration, shown as a light shading in the plot.
The sensitivity has been here separately estimated for the IceCube \textit{upward}, IceCube \textit{contained}, and DeepCore contained events, 
following the procedure described in detail, e.g., in Refs.\cite{Barger:2010ng,Roszkowski:2012uf}.
We take the effective area for IceCube and the effective volume for DeepCore as parametrised in\cite{halzen,Barger:2010ng,Barger:2011em}.
The light shaded points in \reffig{fig:comple1}\subref{fig:b} fall within the reach of at least one of these three estimates. 
The projected 2016 reach in \sigsdp\ at ANTARES\cite{VBertin} in 
the $b\bar{b}$ (green), $W^+W^-$ (purple), and $\tau^+\tau^-$ (cyan) final states is also shown.

\begin{figure}[t]
\centering
\subfloat[]{%
\label{fig:a}%
\includegraphics[width=0.45\textwidth]{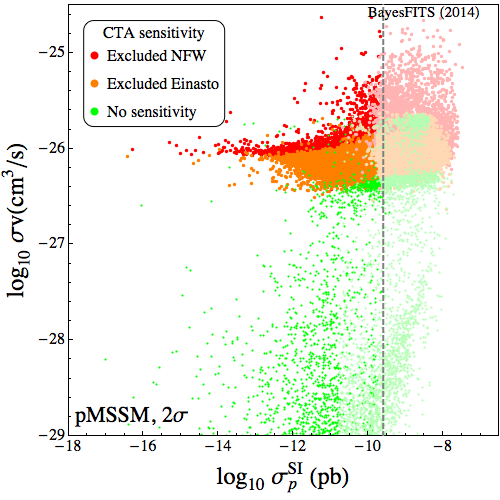}
}%
\hspace{0.07\textwidth}
\subfloat[]{%
\label{fig:b}%
\includegraphics[width=0.45\textwidth]{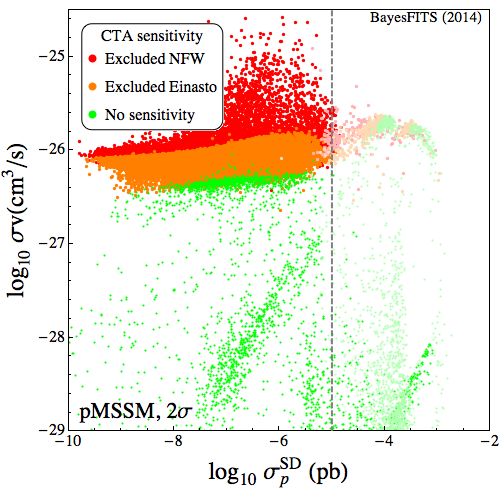}
}%
\caption{\footnotesize In all panels the colour code is the same as in~\reffig{fig:CTA1}\protect\subref{fig:a}. 
Points are shown with $\Delchisq\leq 5.99$ (see Table~\ref{tab:exp_constraints}).
\protect\subref{fig:a} The sensitivity of CTA to the pMSSM points in the (\sigsip, \sigv) plane. 
Lighter shaded points are within the projected sensitivity of 1-tonne detectors. 
The dashed grey line gives an approximate reference value for future direct detection reach in \sigsip.
\protect\subref{fig:b} The sensitivity of CTA to the pMSSM points in the (\sigsdp, \sigv) plane. 
Lighter shaded points are within the projected 5-year sensitivity of IceCube/DeepCore. 
The dashed grey line gives an approximate reference value for future reach in \sigsdp.}  
\label{fig:comple2}
\end{figure}

\begin{figure}[t]
\centering
\subfloat[]{%
\label{fig:a}%
\includegraphics[width=0.45\textwidth]{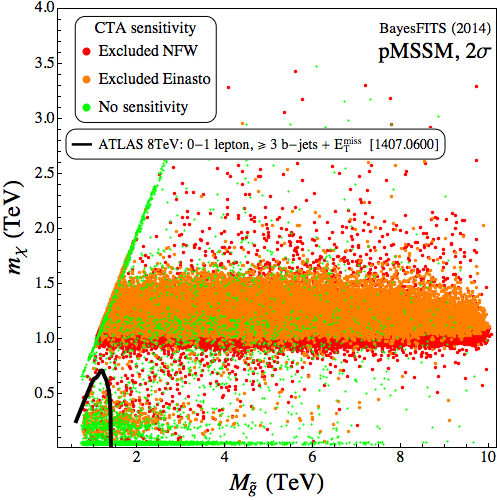}
}%
\hspace{0.07\textwidth}
\subfloat[]{%
\label{fig:b}%
\includegraphics[width=0.45\textwidth]{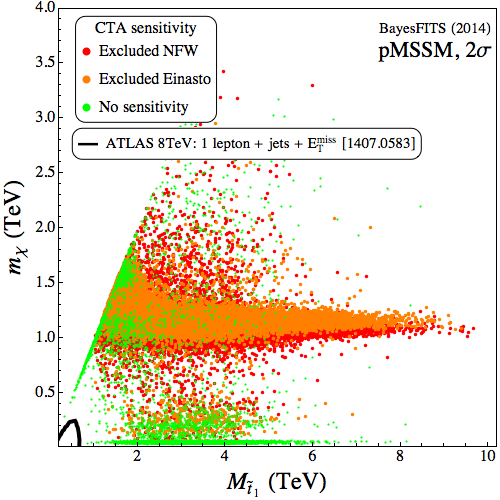}
}%

\subfloat[]{%
\label{fig:c}%
\includegraphics[width=0.45\textwidth]{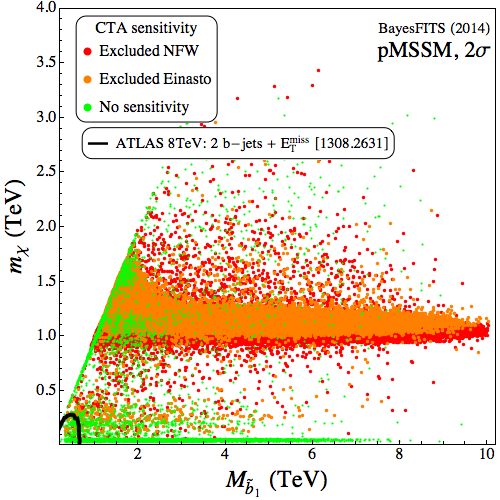}
}%
\hspace{0.07\textwidth}
\subfloat[]{%
\label{fig:d}%
\includegraphics[width=0.45\textwidth]{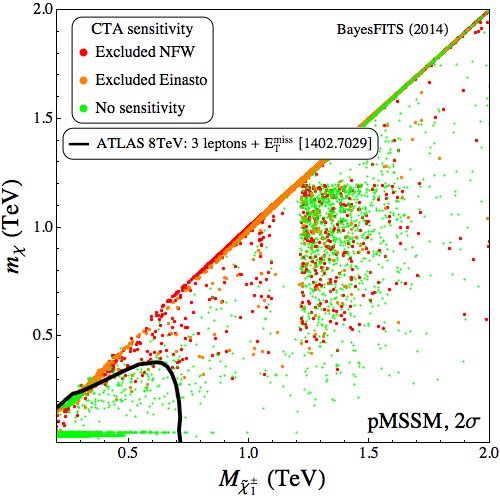}
}%
\caption{\footnotesize In all panels the colour code is the same as in \reffig{fig:CTA1}\protect\subref{fig:a}. 
\protect\subref{fig:a} Sensitivity of CTA in the (\mglu, \mchi) plane, the thick black line shows the current ATLAS limit in the simplified model of Ref.\cite{Aad:2014lra}. 
\protect\subref{fig:b} Sensitivity of CTA in the (\mstopone, \mchi) plane, the thick black line shows the current ATLAS limit in the simplified model of Ref.\cite{Aad:2014kra}.
\protect\subref{fig:c} Sensitivity of CTA in the ($m_{\tilde{b}_1}$, \mchi) plane, 
the thick black line shows the current ATLAS limit in the simplified model of Ref.\cite{Aad:2013ija}.
\protect\subref{fig:d} Sensitivity of CTA in the ($m_{\charone}$, \mchi) plane, 
the thick black line shows the current ATLAS limit in the simplified model of Ref.\cite{Aad:2014nua}.}  
\label{fig:LHC}
\end{figure}

To highlight the idea of complementarity, we show in \reffig{fig:comple2}\subref{fig:a} the reach of CTA compared 
to the one of 1-tonne detectors in the (\sigsip, \sigv) plane. 
The projected sensitivity of CTA appears as horizontal bands that follow the same colour code as in~\reffig{fig:CTA1}\subref{fig:a}, 
while the points within the projected reach of 1-tonne detectors are shown as lighter shaded. 
The dashed grey line vertically separates the points within the sensitivity of 1-tonne detectors (lighter shading) 
from the ones below sensitivity (regular colouring). 
In~\reffig{fig:comple2}\subref{fig:b} we present the equivalent picture in the (\sigsdp, \sigv) plane, 
where the light-shaded region indicates the points within 5-year sensitivity at IceCube/DeepCore, and the dashed grey line separates the points 
within sensitivity from those outside the reach.  

As was mentioned in Secs.~\ref{sec:methodology} and \ref{sec:pointchar}, a detailed analysis of the LHC reach in the pMSSM is 
beyond the scope of this paper because it interests regions of the parameter space characterised by a low mass neutralino,
orthogonal in some sense to the regions in which CTA is most sensitive.
However, for completeness we show in \reffig{fig:LHC} the reach of CTA compared to the present limits on sparticle masses obtained in simplified models at the LHC.
These are meant to be seen as a generic indication of the present reach of the LHC
and the reader must keep in mind that in complex models a detailed analysis can often produce 
limits that can be either weaker or stronger than the ones under simplified model assumptions\cite{Kowalska:2013ica}.  
  
In \reffig{fig:LHC}\subref{fig:a} we show the reach of CTA in the (\mglu, \mchi) plane. The solid black line shows 
the 95\%~C.L. bound obtained at ATLAS for the simplified model of\cite{Aad:2014lra}. 
The limit obtained at CMS is very similar\cite{Chatrchyan:2013iqa}. 
The reach of CTA in the (\mstopone, \mchi),
($m_{\tilde{b}_1}$, \mchi), and ($m_{\charone}$, \mchi) planes is shown in Figs.~\ref{fig:LHC}\subref{fig:b}, 
\ref{fig:LHC}\subref{fig:c}, and \ref{fig:LHC}\subref{fig:d}, respectively. 
With the exception of the coannihilation bands, shown in \reffig{fig:LHC} as regions of mass degeneracy (e.g. $\mchi\approx\mglu$ in \reffig{fig:LHC}\subref{fig:a}),
the reach of CTA is largely independent of the sparticle spectrum, 
as was to be expected. 
Improvements in the limits on the gluino and squark masses are not expected to have any effect on the sensitivity of CTA. 
Indeed, CTA remains sensitive to spectra where the gluinos and squarks lie well beyond the reach of present and future colliders. 

Note that in \reffig{fig:LHC}\subref{fig:d} we have indicated with a solid black line the bound from 3 lepton
EW-ino searches obtained under the assumption that a light slepton of the first two generations is present and has a mass 
in between \mchi\ and $m_{\charone}$ or $m_{\neuttwo}$\cite{Aad:2014nua,Khachatryan:2014qwa}. 
As it stands the limits seems to exclude almost entirely the $Z/h$-resonance region of the parameter space, 
shown in the lower left-hand side of the plot, at $\mchi<100\gev$.  
One must remember that for the points of the pMSSM characterised by heavier 
sleptons the limit is actually weaker than shown here\cite{Aad:2014nua,Khachatryan:2014mma}, 
thus allowing part of the $Z/h$-resonance region to survive the bounds.

\section{Summary and conclusions\label{sec:summary}}

In this paper we have investigated the prospects for detection of neutralino DM in the framework of the pMSSM. 
We have focused here particularly on models for which the LSP is a neutralino at about the 
\tev\ scale, outside the reach of the 14\tev\ run at LHC. 
TeV-scale neutralinos have been shown to be promising candidates for DM in many well-motivated SUSY 
scenarios after the Higgs boson discovery. 

In order to satisfy the constraints from the relic density neutralinos in this mass range must preferentially be nearly pure 
higgsinos, or winos, or a mixture of the two. However, as is known from many studies recently appeared in the literature, 
thermal-relic winos are subject to large Sommerfeld enhancement of their annihilation cross section,
so that they are excluded at the $2\sigma$ level by the H.E.S.S. search for $\gamma$-ray lines from the GC 
under most choices of the DM halo profile.
Consequently the most likely candidate for heavy SUSY DM that is not excluded by the present constraints 
is the $\sim 1\tev$ nearly pure higgsino.

We showed in this paper that the air Cherenkov radiation telescope array of imminent construction CTA 
has the sensitivity reach to nearly exclude at the 95\%~C.L. the $\sim 1\tev$ higgsino region of the pMSSM with 500 hours of observation.
We derived the sensitivity of CTA to continuous and $\gamma$-ray line photons 
arising from neutralino DM annihilation by constructing an energy-binned likelihood function
and adopted the most up to date estimates of the detector response functions and 
modelling of the background provided by the CTA Collaboration. 
We obtained results for the Einasto and NFW profile assumptions.
We applied our results to the parameter space of the pMSSM, but also presented the limits for single annihilation 
final state channels. 

We showed that under the NFW profile assumption 
CTA will be able to probe 70\% of the points belonging to the $\sim 1\tev$ higgsino region in our scans, 
specifically those for which the annihilation cross section is enhanced by one of the following factors: 
resonance with a heavy Higgs boson, $A/H$; 
non-negligible admixture with the wino; or a significant annihilation to a monochromatic $\gamma\gamma$ line.
Assumption of the Einasto profile allows us to formulate even more optimistic projections, 
as we found that in this case CTA will also probe the remaining part of the $\sim1\tev$ higgsino region, with the exclusion of 
some points characterised by heavy squark or gluino co-annihilation, and will probe additionally 
a significant part of the parameter space leading to bino and mixed bino/higgsino DM.

We also found that, in complementarity with other direct and indirect detection experiments, CTA will significantly 
probe the favoured parameter region of the pMSSM, far beyond the 
reach of 1-tonne underground detectors alone. 
We showed that many of the points well within our calculated sensitivity of CTA lie below the onset 
of the irreducible atmospheric and diffuse supernova neutrino background for direct detection.  

Finally we found that by combining different experiments 
detection prospects for SUSY DM are very good also in scenarios where the neutralino is not the only particle comprising
the DM in the Universe. The existing bounds and projections strongly depend in this case on the 
adopted prescription for rescaling the local DM abundance.
We found that even under the most conservative assumption, that the local density is rescaled with the square of the ratio
of the neutralino density to the PLANCK value,  
CTA will be able to exclude a large region of the parameter space, 
significantly outperforming the reach of current Fermi-LAT dSphs and H.E.S.S. $\gamma$-ray line searches. 
\bigskip

\noindent \textbf{Acknowledgments}
\medskip

\noindent  We would like to thank John Carr for useful discussions on the CTA project and for supplying us the 
  latest detector response functions and background simulation. We would also like to thank Vincent Bertin 
  for providing us the projected sensitivity of ANTARES.
  A.J.W. would like to thank Alexander Pukhov
  for clarifying some issues regarding \texttt{micrOMEGAs}. 
  This work has been funded in part by the Welcome Programme
  of the Foundation for Polish Science.
  L.R. is also supported in part by a STFC
  consortium grant of Lancaster, Manchester, and Sheffield Universities. The use
  of the CIS computer cluster at the National Centre for Nuclear Research is gratefully acknowledged. 
L.R. would like to thank the Theory Division of CERN for hospitality during the final stages of the project.
\bigskip
 
\appendix
\section{Derivation of the CTA reach\label{sec:app}}

In this appendix we present the calculation of CTA's sensitivity to the WIMP annihilation cross section.
We compare the sensitivities that can be obtained through different likelihood functions, and we refer to some 
other recent work on the subject that can be found in the literature\cite{Doro:2012xx,Wood:2013taa,Pierre:2014tra,Silverwood:2014yza}.

\begin{figure}[t]
\centering
\includegraphics[width=0.50\textwidth]{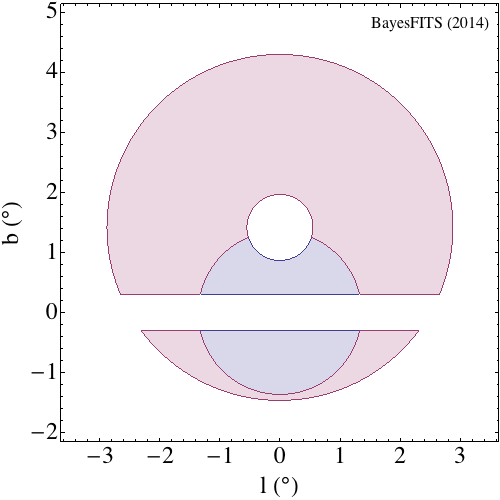}
\caption{\footnotesize In blue, the \textbf{ON} region: the angular radius is $\Delta_{\textrm{cut}}=1.36^{\circ}$.
In pink, the \textbf{OFF} region: the offset from the GC is $b_{\textrm{off}}=1.42^{\circ}$; the inner radius is 
$r_1=0.55^{\circ}$ and the outer radius is $r_2=2.88^{\circ}$. 
The strip of sky at $|b|<0.3^{\circ}$ is subtracted from the ON and OFF regions.}  
\label{fig:regions}
\end{figure}

We use the \textit{Ring Method} as defined in\cite{Doro:2012xx,Wood:2013taa,Pierre:2014tra,Silverwood:2014yza}.
Two regions are identified in the plane of the galactic coordinates $l$ and $b$, as shown in \reffig{fig:regions}.
The ``signal", or ON, region is based on a circle of angular radius $\Delta_{\textrm{cut}}=1.36^{\circ}$ around the GC.
The ``background", or OFF, region is based on a ring centred at the offset coordinate $b_{\textrm{off}}=1.42^{\circ}$,
with an inner angular radius of $r_1=0.55^{\circ}$ and an outer radius of $r_2=2.88^{\circ}$, from which the ON region is subtracted. 
The strip of sky characterised by $|b|<0.3^{\circ}$ about the GC and 
the region of the sky within the inner radius $r_1$ do not belong to either the ON or OFF regions.    


We proceed to create a binned likelihood function. For each energy bin, $i$, the expected number of counts from DM annihilation is calculated:
\begin{equation}
N_i^{\textrm{ann}} = t_{\text{obs}}\cdot J\cdot\frac{\sigv}{8 \pi m^2_\chi} \int_{\Delta E_i}dE\left(\frac{1}{\sqrt{2 \pi \delta(E)^2}}\int_{26\gev}^{\mchi} d\bar{E} \frac{dN_\gamma(\bar{E})}{d\bar{E}} A_{\text{eff}}(\bar{E})e^{-\frac{(E - \bar{E})^2}{2\delta(E)^2}}\right)\,,\label{count}
\end{equation}
where $A_{\text{eff}}$ is the effective area of the detector, $\delta(E)$ is the energy resolution, $dN_{\gamma}/dE$ is the annihilation spectrum, 
and $J$ is the $J$-factor for either the ON or OFF region.
For $A_{\text{eff}}$ and $\delta(E)$ we use the most up to date instrument response functions provided by the CTA Collaboration\cite{JCarr}. 
We take an observation time $t_{\text{obs}} = 500\textrm{ h}$\,.

As is well known, the $J$-factor is defined by integration along the line of sight (l.o.s.) of the DM halo profile squared,
$\rho^2(r)$:
\begin{equation}
J=\int_{\Delta\Omega} \int_{\textrm{l.o.s.}} 
\rho^2[r(\theta)]dr(\theta)d\Omega\,,
\end{equation}
where in the reference frame centred at the observer the distance from the GC depends on the azimuthal angle $\theta$. 
Using the Einasto DM density profile\cite{Einasto}, we obtain $J^{\textrm{Ein}}_{\textrm{ON}}=7.44\times 10^{21}\gev^2/\textrm{cm}^5$ and $J^{\textrm{Ein}}_{\textrm{OFF}}=1.21\times 10^{22}\gev^2/\textrm{cm}^5$ for the $J$-factors of the ON and OFF regions, respectively, in agreement with the results of\cite{Silverwood:2014yza}.
The corresponding values for the NFW\cite{Navarro:1995iw} profile are $J^{\textrm{NFW}}_{\textrm{ON}}=3.89\times 10^{21}\gev^2/\textrm{cm}^5$ and $J^{\textrm{NFW}}_{\textrm{OFF}}=5.78\times 10^{21}\gev^2/\textrm{cm}^5$\,.


To gauge the constraining power of the binned likelihood 
we compare the projected sensitivity for DM annihilating to $b\bar{b}$ for three different methods.\medskip 

1. We first calculate the binned Poisson likelihood defined by
\begin{equation}
{\cal{L}} = \prod_{i,j} \frac{ \mu_{ij}^{n_{ij}}}{n_{ij}!} e^{-\mu_{ij}}\,.\label{binlike}
\end{equation}
In Eq.~(\ref{binlike}) $\mu_{ij}$ is the expected number of photons in bin $ij$, with the index $i$ running over the energy bins and $j=1,2$ for the ON and OFF region, respectively. It is given by the sum of the background and the expected count from DM annihilation,
Eq.~(\ref{count}). $n_{ij}$ is the number of photons counted in bin $ij$.

We find the 95\%~C.L. limit by setting $n_{ij}$ equal to the number of background-only photons 
and calculating ${\cal L}$ by increasing the annihilation cross section from the best fit value (\sigv = 0) 
until the difference in $-2 \ln {\cal L}$ from the best fit value is 2.71 (one-sided 95\%~C.L.).
We use an updated\cite{JCarr} background estimate provided by the CTA Collaboration\cite{montecarlo}.\medskip

2. In the limit of a single energy bin one can construct a test statistics for the
background-only hypothesis\cite{Li:1983fv} 
\begin{equation}
-2\ln{\cal L} = 2\left[N_{\textrm{ON}}  \ln\left(\frac{1 + \alpha}{\alpha} \frac{N_{\textrm{ON}}}{N_{\textrm{ON}} + N_{\textrm{OFF}}}\right) + N_{\textrm{OFF}}  \ln\left((1 + \alpha) \frac{N_{\textrm{OFF}}}{N_{\textrm{ON}} + N_{\textrm{OFF}}}\right) \right]\,,\label{LiMa}
\end{equation}
where $N_{\textrm{ON}}$ and $N_{\textrm{OFF}}$ are the number of counts in the ON and OFF regions, respectively,
and $\alpha= \Delta\Omega_{\textrm{ON}}/\Delta\Omega_{\textrm{OFF}}$ is the ratio of the ON and OFF solid angles.
We calculate the projected sensitivity by setting  $N_{\textrm{ON}}$ and $N_{\textrm{OFF}}$ equal to the expected number of counts
$N_j=\sum_i\mu_{ij}$. We calculate $\alpha= 0.2457$.
We will compare the sensitivity obtained with Eq.~(\ref{LiMa}), 
hereafter referred to as the Li and Ma method,  with the binned likelihood method, Eq.~(\ref{binlike}).\medskip 

3. We finally compare the binned Poisson likelihood, Eq.~(\ref{binlike}), to the 
binned Skellam distribution\cite{Skellam46}, on which the results of\cite{Pierre:2014tra} are based.
Note that the results of Ref.\cite{Pierre:2014tra} were used in our previous study\cite{Roszkowski:2014wqa} 
to estimate the sensitivity of CTA to the parameter space of the CMSSM and the NUHM.

If one calculates for each energy bin $i$ the difference in the measured number of events 
$\theta_{\text{diff},i} = n_{i,\textrm{ON}} - \alpha\, n_{i,\textrm{OFF}}$, 
the probability of observing a set of particular values $\{\theta_{\textrm{diff},i}\}$ is given by the Skellam distribution
\begin{equation}
{\cal L}(\{\theta_{\textrm{diff},i}\}) = \prod_{i} e^{-(N_{i,\textrm{ON}} + \alpha N_{i,\textrm{OFF}})} \left(\frac{N_{i,\textrm{ON}}}{\alpha N_{i,\textrm{OFF}}}\right)^\frac{\theta_{\text{diff},i}}{2} I_{|\theta_{\text{diff},i}|}(2\sqrt{\alpha N_{i,\textrm{ON}} N_{i,\textrm{OFF}}})\,,
\end{equation}
where $N_{i,\textrm{ON}}=\mu_{i1}$, $N_{i,\textrm{OFF}}=\mu_{i2}$ are the expected numbers of photons 
and $I_{|\theta_{\text{diff},i}|}$ is the $|\theta_{\text{diff},i}|$th Bessel function of the first kind. 
The projected limit is obtained by setting $\theta_{\text{diff},i} = 0$ for all the energy bins.  
One finds
\begin{equation}
{\cal L} =  \prod_{i} e^{-(N_{i,\textrm{ON}} + \alpha N_{i,\textrm{OFF}})}I_{0}(2\sqrt{\alpha N_{i,\textrm{ON}} N_{i,\textrm{OFF}}})\,.
\end{equation}
\medskip 

\begin{figure}[t]
\centering
\includegraphics[width=0.50\textwidth]{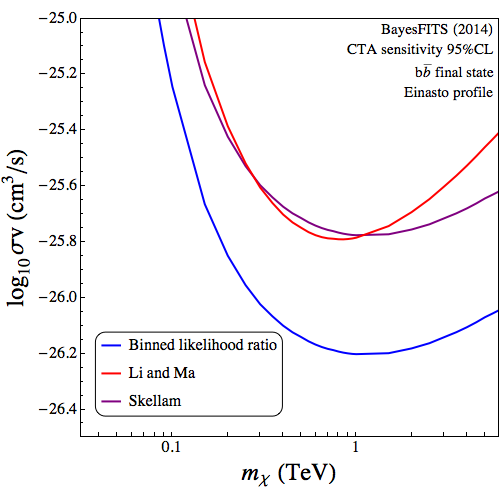}
\caption{\footnotesize Comparison of the limits obtained using the 
three methods of calculating the likelihood for annihilation to the $b\bar{b}$ final state assuming the Einasto profile.}  
\label{fig:p16}
\end{figure}

In \reffig{fig:p16} we show the limits obtained for the $b\bar{b}$ final state and Einasto profile using these three methods.
For low masses the limits obtained using the Skellam distribution and the Li and Ma method are comparable.
For larger masses the Li and Ma method loses sensitivity compared to the other two methods as they benefit from the data being binned by energy.
The binned likelihood method produces the strongest limits and we adopt this method for producing our projected sensitivities.
We find limits that are stronger than the estimates of\cite{Pierre:2014tra} and\cite{Silverwood:2014yza} but comparable to\cite{Wood:2013taa}. 

It should be noted that we have not included uncertainties in the DM distribution, 
systematic uncertainties in the detector response, or included the diffuse $\gamma$-ray background. 
The uncertainty in the DM distribution enters into the calculation of the limits through the $J$-factors in Eq.~(\ref{count}). 
We account for this by presenting limits for two different realisations of the DM distribution, the NFW and Einasto profiles.
We do not consider cored solutions\cite{Salucci:2007tm,Nesti:2013uwa} for the halo profile, 
as they also require a reinterpretation of current limits regarding the wino.
Nevertheless we comment on the impact of the Burkert\cite{Burkert:1995yz} profile assumption on our plots in \refsec{sec:CTA}.

The systematic uncertainty due to the finite energy resolution of the experiment already appears in Eq.~(\ref{count}).
However further systematic effects can be present such as varying acceptance across the field of view or uncertainties in the effective area.

Finally, the diffuse astrophysical $\gamma$-ray background around the galactic centre measured by H.E.S.S.\cite{Aharonian:2006au} 
and Fermi-LAT\cite{Ackermann:2012} presents a challenge to this type of ON/OFF analysis
since this background will be larger in the ON (signal) region than the OFF (background) region mimicking the searched for signal.
Thus the sensitivity presented here is somewhat optimistic and would be reduced due to the diffuse background and systematic uncertainties, 
however the treatment of Ref.\cite{Silverwood:2014yza} suggests that a morphological analysis could partially mitigate these effects.

\begin{figure}[t]
\centering
\includegraphics[width=0.50\textwidth]{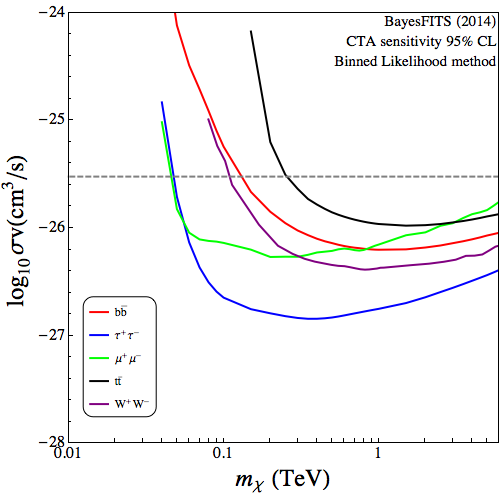}
\caption{\footnotesize 95\%~C.L. CTA projected limits derived for the specific final states most commonly found in the pMSSM with the binned likelihood of Eq.~(\ref{binlike}) for the Einasto profile.
The limits for the $W^+W^-$, $b\bar{b}$, $t\bar{t}$, and $\tau^+\tau^-$ final states are compared to realistic MSSM models in \reffig{fig:finalstate}.}  
\label{fig:p13}
\end{figure}

Figure~\ref{fig:p13} shows the derived 95\%~C.L. limits for some specific final states including the most common primary annihilation 
channels found in the pMSSM using the binned likelihood of Eq.~(\ref{binlike}). 
The limits obtained can probe values of \sigv\ below the ``canonical" thermal relic value 
for all of the final states. 
We compare these to the actual final states found by the scan in \reffig{fig:finalstate}, \refsec{sec:pointchar}.

\bibliographystyle{JHEP}

\bibliography{p19mssm}

\end{document}